\documentclass[a4paper, aps, amsfonts, amssymb, amsmath, prd, showkeys, nofootinbib, twocolumn, superscriptaddress, colorlinks, linkcolor=refcol, citecolor=refcol, urlcolor=refcol,preprintnumbers]{revtex4-2}
\usepackage[english]{babel}
\usepackage[utf8]{inputenc}
\usepackage{orcidlink}
\usepackage{bbold}
\usepackage{physics}
\usepackage{makecell}
\definecolor{refcol}{RGB}{178,34,34}
\usepackage{microtype}
\usepackage{tabularx}
\usepackage{comment}
\newcommand{\be}{\begin{equation}}
\newcommand{\ee}{\end{equation}}

\newcommand{\CP}{\mathrm{CP}}
\newcommand{\cl}{\mathrm{cl}}
\newcommand{\vac}{\mathrm{vac}}
\newcommand{\mat}{\mathrm{mat}}
\newcommand{\eff}{\mathrm{eff}}

\definecolor{red}{rgb}{1,0,0}
 
\definecolor{internationalkleinblue}{rgb}{0.0, 0.18, 0.65}

\allowdisplaybreaks

\begin{document}
\title{Lee-Yang zeros and edge singularity in a mean-field approach}

\author{Tatsuya Wada \orcidlink{0009-0007-4809-5642}}
\email{tatsuya.wada@yukawa.kyoto-u.ac.jp}
    \affiliation{Yukawa Institute for Theoretical Physics, Kyoto University, Kyoto, 606-8502, Japan}

\author{Győző Kovács \orcidlink{0000-0003-0570-3621}}
\email{gyozo.kovacs@uwr.edu.pl}
    \affiliation{Institute of Theoretical Physics, University of Wroc\l{}aw, plac Maksa Borna 9, PL-50204 Wroc\l{}aw, Poland}
    \affiliation{Institute for Particle and Nuclear Physics,
    HUN-REN Wigner Research Centre for Physics, 1121 Budapest, Konkoly–Thege Miklós út 29-33, Hungary}
    
\author{Masakiyo Kitazawa \orcidlink{0000-0002-5210-7649}}
\email{kitazawa@yukawa.kyoto-u.ac.jp}
    \affiliation{Yukawa Institute for Theoretical Physics, Kyoto University, Kyoto, 606-8502, Japan}
    \affiliation{J-PARC Branch, KEK Theory Center, Institute of Particle and Nuclear Studies, KEK, Tokai, Ibaraki 319-1106, Japan}

\author{Takahiro M. Doi \orcidlink{0000-0003-0367-1312}}
\email{doi.takahiro.5d@kyoto-u.ac.jp}
    \affiliation{Division of Physics and Astronomy, Graduate School of Science, Kyoto University, Kitashirakawaoiwake, Sakyo, Kyoto 606-8502, Japan}

\begin{abstract}
The analytic structure of the partition function in finite-volume systems is investigated at complex chemical potentials in a minimal mean-field effective model of QCD with finite-size effects incorporated. 
We discuss the temperature dependence of the Lee-Yang zeros and their relation to the edge singularity for various system sizes.
Different methods for locating the critical point based on finite-size scaling of Lee-Yang zeros and susceptibility ratios are compared. 
We demonstrate that these methods can successfully identify the critical point, whereas a careful treatment of corrections from irrelevant operators is crucial for its accurate determination.
\end{abstract}

\preprint{YITP-26-59,J-PARC-TH-0336}
\date{\today}

\hypersetup{
pdftitle={Lee-Yang zeros and edge singularity in a mean-field approach},
pdfauthor={Wada Tatsuya, Kovacs Gyozo, Kitazawa Masakiyo, Doi Takahiro}
}

\maketitle

\section{Introduction} \label{sec:intro}

Investigation of the critical behavior associated with phase transitions is of main interest not only in quantum chromodynamics (QCD), but also in various physical systems. However, locating a second-order critical point (CP) using thermodynamic observables in finite-volume systems is a nontrivial task. While the universal divergence of susceptibilities uniquely marks the critical region in the thermodynamic limit, in a finite system such divergence is always smeared, and all thermodynamic quantities become analytic~\cite{Cardy:1988ag, Binder:1992pz}.
To locate the CP from the information at finite volumes, finite-size scaling (FSS) offers a systematic approach~\cite{Fisher:1972zza,Binder:1992pz,Privman:1983,Pelissetto:2000ek,Binder:2001ha}. Such analyses are particularly important in numerical simulations, because they are inevitably performed in finite-volume systems. Various methods to locate the CP using susceptibilities and their ratios based on FSS have been developed and applied to various systems in the literature~\cite{Pelissetto:2000ek,Binder:2001ha,Karsch:2000xv,Jin:2017jjp,Ferrenberg:2018zst,Kuramashi:2020meg,Cuteri:2020yke,Cuteri:2021ikv,Philipsen:2021qji,Kiyohara:2021smr,Ashikawa:2024njc,Ejiri:2026ijj}.

An interesting feature of finite-size thermal systems is that the partition function has discrete zeros, known as the Lee-Yang (or Fisher) zeros~\cite{Yang:1952be,Lee:1952ig,Fischer:1965rna}, in the complexified parameter space. The distribution of the Lee-Yang zeros (LYZs) is known to encode information of phase transitions~\cite{Yang:1952be,Lee:1952ig,Itzykson:1983gb,Bena:2005}. In the thermodynamic limit, the LYZs accumulate densely along a line in the complex-parameter plane, forming a branch cut, and their endpoint, called the Lee-Yang edge singularity (LYES)~\cite{Fisher:1978pf}, reaches the real axis at the CP. The LYZs and LYES have been investigated extensively in the literature~\cite{Halasz:1996jg, Ejiri:2005ts, Stephanov:2006dn, Ejiri:2014oka, An:2017brc, Giordano:2019gev, Wakayama:2019hgz, Connelly:2020gwa, Mukherjee:2021tyg, Rennecke:2022ohx, Li:2022dqq, Li:2024ymo, Zhang:2025jyv, Wan:2025wdg,DiRenzo:2025nya,Skokov:2024fac,DiRenzo:2024izy,Karsch:2023rfb,Wada:2024qsk, Wada:2025ycz}. Recently, the use of LYZs and LYES in the search for the CP has also been actively explored in lattice QCD simulations~\cite{Dimopoulos:2021vrk, Bollweg:2022rps, Basar:2023nkp, Clarke:2024ugt,
Schmidt:2025ppy,Adam:2025phc,Borsanyi:2025ttb,Kitazawa:2025cdg}.

In order to gain a better understanding of the behavior of LYZs in QCD, investigations based on effective models can provide valuable insights, provided that finite-size effects are properly incorporated. The finite-size effects in effective models of QCD have been discussed in the literature within the mean-field approach in the linear sigma model (LSM)~\cite{Palhares:2009tf, Magdy:2015eda, Kovacs:2023kbv} and the Nambu--Jona-Lasinio (NJL) model~\cite{Bhattacharyya:2012rp, Pan:2016ecs, Wang:2018qyq}, and also using functional methods, e.g., in functional renormalization group (FRG) \cite{Tripolt:2013zfa, Almasi:2016zqf} and Dyson-Schwinger (DS) approaches \cite{Bernhardt:2021iql, Bernhardt:2022mnx}. In mean-field studies, the finite-size effects have been taken into account in the effective potential $\mathcal{U}(\bar\phi)$ for a uniform field, such as the chiral order parameter, $\bar\phi$ through the momentum-space discretization imposed by the boundary conditions. The mean-field value of $\bar\phi$ is then determined by the minimization of $\mathcal{U}(\bar\phi)$, and the minimum value defines the free energy. 
However, this approach yields non-analytic phase transitions even for finite-size systems, which contradicts the analyticity of the partition function in finite systems and FSS. Also, this approach cannot describe LYZs, because they emerge through cancellations of contributions from different field configurations in the partition function. Within functional methods, the finite-size rounding at the CP can be investigated; however, to study the behavior of LYZs, a careful treatment at finite sizes in the complex parameter space would be necessary, which raises significant difficulties and has not yet been fully performed. 

These issues in the mean-field approaches can be resolved in a minimal extension by keeping the field $\bar\phi$ spatially uniform, but incorporating its fluctuations in the partition function by an integration over $\bar\phi$~\cite{Kovacs:2025gct}. As a result, the partition function becomes analytic in finite volume, while the conventional mean-field approach is recovered in the thermodynamic limit. Integration over the constant mode can be viewed as a zero-dimensional field theory \cite{Laine:2016hma}, while an analogous treatment of the zero mode is used in chiral perturbation theory to describe the $\epsilon$-regime~\cite{Gasser:1986vb, Gasser:1987ah}.
In Ref.~\cite{Kovacs:2025gct}, this method has been introduced for the analysis of susceptibilities in the quark-meson model at nonzero temperature ($T$) and baryon chemical potential ($\mu_{\rm B}$). It has been shown that the susceptibilities at finite volume behave smoothly even in the vicinity of the CP and the first-order phase transition line. Their behavior near the CP is consistent with FSS with the mean-field critical exponents. 

In the present study, we apply the framework of Ref.~\cite{Kovacs:2025gct} to explore the behavior of the LYZs in the complex $\mu_{\rm B}$ plane within the same quark-meson model. Since this approach reproduces the conventional mean-field phase diagram in the thermodynamic limit, it allows a simultaneous analysis of both the LYZs in finite volume and the LYES in the infinite-volume limit~\cite{Mukherjee:2021tyg, Zhang:2025jyv} in an effective model. We explicitly demonstrate the system-size dependence of the LYZs and their approach to the LYES. In addition to consistency with FSS, we argue that the system-size dependence of the LYZs is more naturally understood in the complex $\mu_{\rm B}^2$ plane than in the complex $\mu_{\rm B}$ plane. We compare our findings qualitatively with the lattice QCD results for the LYZs~\cite{Basar:2023nkp, Clarke:2024ugt, Adam:2025phc}. 

We also test methods for locating the CP using numerical results obtained at finite volume. We compare the methods using intersection points based on the LYZ ratio (LYZR)~\cite{Wada:2024qsk,Wada:2025ycz} and the Binder-cumulant~\cite{Binder:1981sa}. We demonstrate that these methods successfully determine the location of the CP within $1\%$ accuracy in the quark-meson model. However, we find that the convergence of the intersection point in the LYZR method toward the thermodynamic limit becomes slow beyond this level of accuracy. We argue that this behavior is attributed to corrections to FSS arising from irrelevant operators. 

This paper is organized as follows. In Sec.~\ref{sec:framework} we introduce the method to treat finite-size systems proposed in Ref.~\cite{Kovacs:2025gct}, and show that the scaling behavior in this framework is consistent with that in the mean-field models.
Our model choice and its phase diagram in the thermodynamic limit are discussed in Sec.~\ref{sec:model}. Sec.~\ref{sec:LYZandES} is devoted to the discussion of the LYZs and LYES in our model framework. With these at hand, we investigate different methods to locate the CP from the information in finite-size systems in Sec.~\ref{sec:CP}. Finally, in Sec.~\ref{sec:Conclusion} we give our conclusion and outlook. Appendix~\ref{sec:irrelevant} is included to discuss scaling violations due to irrelevant operators.

\section{Description of Finite-size Effects} \label{sec:framework}

\subsection{Finite-volume mean-field approach} \label{sec:finV_MF}

To describe the finite-size effects in a manner consistent with the mean-field approximation, we follow the method discussed in Ref.~\cite{Kovacs:2025gct}.

Near a CP, properties of a thermal system are well characterized by an order-parameter field. Assuming that it is a real scalar field denoted by $\phi(x)$ and integrating out all degrees of freedom except for $\phi(x)$, the partition function is given by 
\begin{align}
    \mathcal{Z} = \int \mathcal{D}\phi e^{-S_{\rm E}(\phi)} \,,
    \label{eq:Zexact}
\end{align}
with the effective Euclidean action $S_{\rm E}(\phi)$. 
In the conventional mean-field approximation, one introduces the effective potential $\mathcal{U}(\bar\phi)=(T/V)S_\mathrm{E}(\bar\phi)$ assuming that $\phi(x)$ is temporally and spatially uniform, $\bar\phi=\phi(x)$, where $T$ and $V$ are the temperature and volume of the system. 
In the standard treatment for the thermodynamic limit, the value of $\bar\phi$ is determined so as to minimize $\mathcal{U}(\bar\phi)$, and the free-energy density is given by $f=\mathcal{U}(\bar\phi)$ at the minimum.
As a result, in this prescription, the value of $\bar\phi$ can have discontinuities or non-analytic behaviors with the variation of thermodynamic parameters, which result in non-analyticities of the free energy corresponding to phase transitions. However, when this procedure is directly applied to a finite-size system, such non-analyticities contradict the fact that the thermodynamics in a finite system is always analytic.

To realize the analyticity at finite $V$ with a minimal modification to the mean-field approximation, we assume that the partition function is given by~\cite{Kovacs:2025gct}
\begin{align}
    \mathcal{Z} = \int d\bar\phi e^{-V\mathcal{U}(\bar\phi)/T} \,.
    \label{eq:ZMF}
\end{align}
Equation~\eqref{eq:ZMF} is regarded as a modification of the mean-field approximation that incorporates fluctuations of the spatially uniform mode of $\phi(x)$. The free-energy density is then defined from Eq.~\eqref{eq:ZMF} as
\begin{align}
f 
=-\frac{T}{V}\ln\mathcal{Z} \,,
\label{eq:f}
\end{align}
which, in contrast to the conventional mean-field approach, does not coincide with $\mathcal{U}(\bar\phi)$. Thermodynamic quantities are obtained from Eq.~\eqref{eq:f} using standard thermodynamic relations. In the thermodynamic limit $V\to\infty$, the conventional mean-field results are recovered because only the global minimum of $\mathcal{U}(\bar\phi)$ contributes to the integral in Eq.~\eqref{eq:ZMF}, while Eq.~\eqref{eq:ZMF} is an analytic function at finite $V$ provided the analyticity of $\mathcal{U}(\bar\phi)$. 

\subsection{Landau potential}
\label{sec:Landau}

To examine how Eq.~\eqref{eq:ZMF} describes thermodynamics near a CP at finite volume, let us employ the Landau potential
\begin{align}
    U_{\rm Landau}(\bar\phi;\tau,\eta) = u_0(\tau,\eta) T + \frac\tau2 \bar\phi^2 + \frac b{4!} \bar\phi^4 - \eta \bar\phi \,,
    \label{eq:Landau}    
\end{align}
where $\tau$ and $\eta$ correspond to the reduced temperature and external magnetic field in the Ising model, respectively. The CP is located at $(\tau,\eta)=(0,0)$ as an endpoint of the first-order phase-transition line at $\eta=0$ and $\tau<0$. The $\bar\phi$-independent term $u_0(\tau,\eta)T$ is assumed to be regular. Substituting Eq.~\eqref{eq:Landau} into Eq.~\eqref{eq:ZMF} yields the partition function 
\begin{align}
    \mathcal{Z}_{\rm Landau}(\tau,\eta,L) = e^{-L^du_0} \int d\bar\phi \, e^{-L^d(\tau \bar\phi^2/2 + b \bar\phi^4 /4! - \eta \bar\phi)/T} \,,
    \label{eq:Z_Landau}
\end{align}
with the system size $L$ and dimensionality $d$.

Performing the scaling transformation
\begin{align}
    \tau\to \ell^{y_\tau} \tau ,
    \quad
    \eta\to \ell^{y_\eta} \eta ,
    \quad
    \bar\phi\to \ell^{y_\phi} \bar\phi ,
    \quad
    L \to \ell^{-1} L ,
    \label{eq:scale_tr}
\end{align}
in Eq.~\eqref{eq:Z_Landau}, one finds that the partition function satisfies the finite-size scaling (FSS) relation
\begin{align}
    &\mathcal{Z}_{\rm Landau}(\tau,\eta,L) 
    \notag \\
    &= \frac{e^{-L^du_0(\tau,\eta)}\ell^{y_\phi}}{e^{-L^du_0(\ell^{y_\tau}\tau,\ell^{y_\eta}\eta)}}  \mathcal{Z}_{\rm Landau}(\ell^{y_\tau}\tau,\ell^{y_\eta} \eta,\ell^{-1} L)
    \notag \\
    &= \frac{e^{-L^du_0(\tau,\eta)}L^{y_\phi}}{e^{-L^du_0(L^{y_\tau}\tau,L^{y_\eta}\eta)}} \tilde{\mathcal{Z}}(L^{y_\tau}\tau,L^{y_\eta}\eta) \,,
    \label{eq:Zscaling}
\end{align}
with the scaling exponents
\begin{align}
    y_\tau = \frac d2 , 
    \quad
    y_\eta = \frac{3d}4 , 
    \quad
    y_\phi = \frac d4 ,
    \label{eq:y}
\end{align}
where on the second equality of Eq.~\eqref{eq:Zscaling} we substituted $\ell=L$ and introduced the scaling function 
\begin{align}
    \tilde{\mathcal{Z}}(\tilde\tau,\tilde\eta) 
    \equiv \int d\phi e^{-(\tilde\tau \bar\phi^2/2 + b \bar\phi^4 /4! - \tilde\eta \bar\phi)/T} \,.    
\end{align}
Equation~\eqref{eq:Zscaling} also leads to the FSS relation for the singular part of the free-energy density
\begin{align}
    f_{\rm s}(\tau,\eta,L) = \tilde{f}_{\rm s}(L^{y_\tau}\tau, L^{y_\eta}\eta ) \,,
    \label{eq:fscaling}
\end{align}
with the scaling function $\tilde{f}_{\rm s}(\tilde\tau, \tilde\eta ) \equiv-T\ln\mathcal{Z}(\tilde\tau, \tilde\eta )$.

When a CP manifests itself in a general system, such as QCD, the variables $\tau$ and $\eta$ are linearly related to thermodynamic parameters in the general system in its vicinity. For the QCD CP on the temperature ($T$)--baryon chemical potential ($\mu_{\rm B}$) plane, for instance, one has
\begin{align}
    \begin{pmatrix} \tau \\ \eta \end{pmatrix} = 
    \begin{pmatrix} a_{11} & a_{12}\\ a_{21} & a_{22}\end{pmatrix}
    \begin{pmatrix} T-T_{\rm CP} \\ \mu_\mathrm{B}-\mu_{\rm CP} \end{pmatrix}
    \equiv A 
    \begin{pmatrix} \delta T \\ \delta \mu_\mathrm{B}\end{pmatrix} ,
    \label{eq:map}    
\end{align}
with the location of the critical point $(T,\mu_{\rm B}) = (T_{\rm CP},\mu_{\rm CP})$.
In this case, the partition function and the singular part of the free-energy density near the CP satisfy
\begin{align}
    \mathcal{Z}(T,\mu_{\rm B},L) &= e^{-L^du_0} L^{y_\phi} \tilde{\mathcal{Z}}\big(L^{y_\tau}\tau(T,\mu_{\rm B}),L^{y_\eta}\eta(T,\mu_{\rm B})\big) \,,
    \label{eq:Zmix} \\
    f_{\rm s}(T,\mu_{\rm B},L) &= \tilde{f}_{\rm s}\big(L^{y_\tau}\tau(T,\mu_{\rm B}),L^{y_\eta}\eta(T,\mu_{\rm B})\big) \,.
    \label{eq:fmix}
\end{align}
The unit vectors along the $\tau$- and $\eta$-axes, $\vec{e}_\tau$ and $\vec{e}_\eta$, are related to those along the $T$- and $\mu_{\rm B}$-axes, $\vec{e}_T$ and $\vec{e}_\mu$, as
\begin{align}
    \vec{e}_\tau \propto a_{22} \vec{e}_T - a_{21} \vec{e}_{\mu_\mathrm{B}}\,, 
    &&
    \vec{e}_\eta \propto a_{12} \vec{e}_T - a_{11} \vec{e}_{\mu_\mathrm{B}}\,.
    \label{eq:axes}
\end{align}

Several remarks are in order. 
First, when the Landau's potential~\eqref{eq:Landau} has higher order terms of $\bar\phi$, such as $\bar\phi^6$, the FSS relations Eqs.~\eqref{eq:Zscaling} and~\eqref{eq:fscaling} are violated due to these terms. However, as shown in Appendix~\ref{sec:irrelevant} the scaling exponents corresponding to these terms are negative, and hence they behave as irrelevant operators whose effects are suppressed for $L\to\infty$. Their influence on the numerical results will be discussed in later sections. 
We also note that the $\bar\phi^6$ term plays the crucial role in describing the tricritical point~\cite{Ejiri:2014oka, Hatta:2002sj}, while it can also improve the description away from the critical region \cite{Braun:2023qak, Mei:2025ein}.
Second, the scaling exponents in Eq.~\eqref{eq:y} correspond to the mean-field values obtained for the Landau potential. Note that a similar form to Eq.~\eqref{eq:Z_Landau} arises in the mean-field approximation of the Ising model, formulated in terms of a continuous magnetization field in leading order \cite{Krasnytska:2016}. When the effects beyond the mean-field approximation are incorporated, it is known that the exponents are modified to the values specific to the universality class. For example, in the three-dimensional Ising model, the values are known as $y_\tau\simeq1.5874$ and $y_\eta\simeq2.4819$ \cite{Rattazzi:2008pe,Chang:2024whx}.
Finally, in the above argument, it is implicitly assumed that $\mathcal{U}(\bar\phi)$ does not have $L$ dependence. However, at finite volume, the momenta become discrete owing to boundary conditions, and the momentum integrals are replaced by a sum over discrete modes, introducing $L$ dependence into $\mathcal{U}(\bar\phi)$~\cite{Palhares:2009tf, Bhattacharyya:2012rp, Magdy:2015eda, Pan:2016ecs, Wang:2018qyq, Kovacs:2023kbv}. Such an effect, however, may be suppressed exponentially because the discretization error of an integral is typically exponentially suppressed for smooth functions~\cite{Trefethen2014}. Accordingly, its impact on the above argument at the polynomial order in $L$ is expected to be negligible.

\subsection{Lee-Yang zeros and Lee-Yang edge singularity}
\label{sec:LYZ}

The partition function of a finite system generally has zero points in the complex-parameter plane. We refer to the zeros of Eq.~\eqref{eq:Z_Landau} in the complex $\eta$ plane, $\eta_{\rm LY}^{(n)}(\tau,L)\in \mathbb{C}$, with real $\tau$ as the Lee-Yang zeros (LYZs)~\cite{Lee:1952ig,Yang:1952be}, i.e., 
\begin{align}
    \mathcal{Z}_{\rm Landau}\big(\tau,\eta_{\rm LY}^{(n)}(\tau,L),L\big)=0 \,,
    \label{eq:LYZ_Z}
\end{align}
where $n$ labels different LYZs on the upper-half plane in the order $0<{\rm Im}\,\eta_{\rm LY}^{(1)}<{\rm Im}\,\eta_{\rm LY}^{(2)}<\cdots$; from Cauchy's reflection theorem, there also exist LYZs at $\eta=\eta_{\rm LY}^{(n)}(\tau,L)^*$ on the lower-half plane. Because of Eq.~\eqref{eq:Zscaling}, by defining the zeros of $\tilde{ \mathcal Z}$ as $\tilde{\mathcal Z}\big(\tilde\tau,\tilde\eta_{\rm LY}^{(n)}(\tilde\tau)\big)=0$, the LYZs satisfy
\begin{align}
    \eta_{\rm LY}^{(n)}(\tau,L) = L^{-y_\eta} \tilde\eta_{\rm LY}^{(n)}(L^{y_\tau}\tau)\,,
    \label{eq:LYZscaling}
\end{align}
provided that the LYZs do not belong to $e^{-L^du_0}$, which is empirically justified near the CP for small $n$~\cite{Wada:2024qsk,Wada:2025ycz}. In accordance with the Lee-Yang's theorem~\cite{Yang:1952be}, Eq.~\eqref{eq:LYZscaling} is pure imaginary. 

For $\tau<0$, reflecting the first-order phase transition at $\eta=0$ the distribution of LYZs around this point becomes denser and the density diverges for $L\to\infty$. On the other hand, for $\tau>0$ the LYZs distribute on the lines starting from complex values $\eta=\pm\eta_{\rm ES}(\tau)$, which are called the Lee-Yang edge singularity (LYES). 
In the thermodynamic limit, the first LYZ approaches the LYES as governed by the FSS relation analogous to Eq.~\eqref{eq:LYZscaling}, with the exponents of the corresponding universality class.
The LYES are obtained by solving the coupled equations \cite{Mukherjee:2021tyg, Zhang:2025jyv}
\begin{align}
    \frac{\partial \mathcal{U}(\bar\phi)}{\partial \bar\phi}=0 \,, 
    \qquad 
    \frac{\partial^2 \mathcal{U}(\bar\phi)}{\partial\bar\phi^2} =0 \,.
    \label{eq:LYES_cond}
\end{align} 
Notice that the solutions of Eq.~\eqref{eq:LYES_cond} are found for complex values of $\bar\phi$ and $\eta$ for $\tau>0$. Since $\mathcal{U}$ is complex in general in this case, Eq.~\eqref{eq:LYES_cond} is a coupled set of four equations for the respective real and imaginary parts. For $\tau<0$, the solutions are found for real $\bar\phi$ and $\eta$, which correspond to the spinodal points. 

For Eq.~\eqref{eq:Landau}, the solutions of Eq.~\eqref{eq:LYES_cond} are found to be
\begin{align}
    \eta_{\rm ES}(\tau) = \pm i z_{\rm ES} \tau^{3/2} \,,
    && z_{\rm ES}=\frac23 \sqrt{\frac2b} \,.
    \label{eq:eta_ES}
\end{align}
In the general system described by Eq.~\eqref{eq:map}, Eq.~\eqref{eq:LYES_cond} gives 
\begin{align}
    ( a_{21} \delta T + a_{22} \delta\mu_{\rm B})^2 = -z_{\rm ES}^2( a_{11} \delta T + a_{12} \delta \mu_{\rm B})^3 \,.
    \label{eq:aES}
\end{align}
Solving Eq.~\eqref{eq:aES} perturbatively in $\delta T$, the solutions connected to $\delta\mu_{\rm B}=0$ at $\delta T=0$ are expanded as
\begin{align}
    \mu_{\rm ES}(T) 
    &= \mu_{\rm CP}- \frac{a_{21}}{a_{22}} \delta T \pm i z_{\rm ES} \frac{(\det A)^{3/2}}{a_{22}^{5/2}} \delta T^{3/2} 
    \notag \\
    &\phantom= + \mathcal{O}(\delta T^2) \,.
    \label{eq:mu_ES}
\end{align}
From Eq.~\eqref{eq:mu_ES}, one finds that the real part of $\mu_{\rm ES}(T)$ with real $T$ passes through $(T_{\rm CP},\mu_{\rm CP})$ and extend along the direction parallel to $\vec{e}_\tau$ in Eq.~\eqref{eq:axes}, while the exponent of $\delta T$ in the imaginary part is the same as that of $\tau$ in Eq.~\eqref{eq:eta_ES}~\cite{Stephanov:2006dn}.

\subsection{Intersection analysis of the critical point}
\label{sec:intersection}

Since the thermodynamics of a finite-size system does not exhibit any singular behavior, locating the CP from the information of finite-size systems is a non-trivial task. In Refs.~\cite{Wada:2024qsk,Wada:2025ycz}, methods to utilize the LYZs for this purpose have been proposed. 
Near the CP, one can Taylor expand $\tilde \eta_{\rm LY}^{(n)}(\tilde\tau)$ as 
\begin{align}
    \tilde \eta_{\rm LY}^{(n)}(\tilde\tau)= i( X_n+Y_n\tilde\tau+\cdots) \,,
    \label{eq:tildeeta}
\end{align}
with $X_n$ and $Y_n$ being real numbers. Substituting it into Eq.~\eqref{eq:LYZscaling} one obtains
\begin{align}
    L^{y_\eta} {\rm Im}\, \eta_{\rm LY}^{(n)}(\tau,L) = X_n + Y_n L^{y_\tau} \tau + \mathcal{O}(\tau^2) \,.
    \label{eq:LYZS}
\end{align}
Equation~\eqref{eq:LYZS} shows that $L^{y_\eta} {\rm Im}\, \eta_{\rm LY}^{(n)}(\tau,L)$ for different values of $L$ intersect at the CP, $\tau=0$, which in turn means that the CP can be determined as the intersection point of Eq.~\eqref{eq:LYZS} if the value of $y_\eta$ is known. Similarly, one finds from Eq.~\eqref{eq:LYZS} that the ratios of the LYZs behave as
\begin{align}
    R_{nm}(\tau,L) 
    &\equiv \frac{\eta_{\rm LY}^{(n)}(\tau,L)}{\eta_{\rm LY}^{(m)}(\tau,L)} 
    \notag \\
    &= r_{nm} + c_{nm} L^{y_\tau} \tau + \mathcal{O}(\tau^2) \,,
    \label{eq:LYZR}
\end{align}
with $r_{nm}=X_n/X_m$, $c_{nm}=r_{nm}\left(Y_n/X_n-Y_m/X_m\right)$. Therefore, the intersection point of Eq.~\eqref{eq:LYZR} is also used for determining the location of the CP. Notable characteristics of Eq.~\eqref{eq:LYZR} compared with Eq.~\eqref{eq:LYZS} are, (1)~the intersection analysis can be performed without knowing the value of $y_\eta$, and (2)~the value of $R_{nm}(\tau,L)$ at the intersection point, $r_{nm}$, is a universal constant specific to the universality class~\cite{Wada:2024qsk}. From the $\tau$ dependence of $R_{nm}(\tau,L)$ around $\tau=0$, one can also examine the value of the critical exponent $y_\tau$.

Another conventional quantity for the intersection analysis is the so-called (fourth-order) Binder cumulant~\cite{Binder:1981sa}, 
\begin{align}
    B_4(\tau,L) 
    \equiv \frac{\langle\bar\phi^4\rangle_{\rm c}}{\langle\bar\phi^2\rangle_{\rm c}^2}\bigg|_{\eta=0} + 3
    = -\frac TV \frac{\partial_\eta^4 f}{(\partial_\eta^2 f)^2}\bigg|_{\eta=0} + 3 \,,
    \label{eq:B4def}
\end{align}
where $\langle \cdot \rangle_{\rm c}$ means the cumulants~\cite{Asakawa:2015ybt} and $\partial_\eta=\partial/\partial(\eta/T)$. Assuming that the free-energy density is dominated by the singular part, Eq.~\eqref{eq:fscaling} yields
\begin{align}
    B_4(\tau,L) = b_4 + c_4 L^{y_\tau} \tau + \mathcal{O}(\tau^2) \,,
    \label{eq:B4}
\end{align}
which again indicates that this quantity obtained on various $L$ can be used for the intersection analysis. Here, $b_4$ is the value of $B_4(\tau,L)$ at the intersection point, which is a universal constant specific to the universality class~\cite{Binder:1981sa}.

The values of $r_{nm}$ and $b_4$ in the mean-field models are determined as follows.
First, $r_{nm}$ are obtained by finding the solutions of 
\begin{align}
    \int_{-\infty}^{\infty} d\bar \phi\, e^{-(\bar\phi^4 -\eta \bar\phi)} =0 \,,
\end{align}
which lie on the imaginary $\eta$ axis, numerically and taking their ratios. This procedure yields
\begin{align}
    r_{21} &\simeq 1.9645, &
    r_{31} &\simeq 2.7902, &
    r_{41} &\simeq 3.5411.
    \label{eq:r_n1_MF}
\end{align}
Second, the MF value of $b_4$ is calculated to be
\begin{align}
    b_4 = \frac{\int_{-\infty}^{\infty} d\bar \phi\, \bar\phi^4 e^{-\bar\phi^4} \int_{-\infty}^{\infty} d\bar \phi\, e^{-\bar\phi^4}}{(\int_{-\infty}^{\infty} d\bar \phi\, \bar\phi^2 e^{-\bar\phi^4})^2}
    \simeq 2.1884 .
    \label{eq:b4_MF}
\end{align}

In the vicinity of the critical point in general systems, by combining Eq.~\eqref{eq:tildeeta} and Eq.~\eqref{eq:map} one obtains~\cite{Wada:2024qsk}
\begin{align}
    \mu_{\rm LY}^{(n)}(T,L) = \mu_{\rm CP} + \frac{i X_n - ( a_{21} L^{y_\eta} -i Y_n a_{11} L^{y_\tau} ) \delta T }{ a_{22} L^{y_\eta} - i Y_n a_{12} L^{y_\tau}} \,.
    \label{eq:LYZmu}
\end{align}
From Eq.~\eqref{eq:LYZmu}, one finds that the behaviors of the LYZs and their ratios are modified as 
\begin{align}
    \mathcal{R}_{nm}(T,L) 
    &\equiv \frac{{\rm Im}\,\mu_{\rm LY}^{(n)}(T,L)}{{\rm Im}\,\mu_{\rm LY}^{(m)}(T,L)}
    \notag \\
    &= \big( r_{nm} + C_{nm} L^{y_\tau} \delta T + \mathcal{O}(\delta T^2) \big)
    \notag \\
    &\phantom{=.} \big( 1 + D_{nm} L^{2\bar y} + \mathcal{O}(L^{4\bar y}) \big) \,,
    \label{eq:LYZRmix} \\
    L^{y_\eta}{\rm Im}\,\mu_{\rm LY}^{(n)}(T,L) &= \bigg( \frac{X_n}{a_{22}} + \frac{Y_n\det A}{a_{22}^2} L^{y_\tau} \delta T + \mathcal{O}(\delta T^2) \bigg)
    \notag \\
    &\phantom{=.} \big(1 + C_n L^{\bar y} + \mathcal{O}(L^{2\bar y} ) \big) \,,
    \label{eq:LYZSmix}
\end{align}
with $\bar y = y_\tau - y_\eta$, which is negative since $y_\tau < y_\eta$ generally holds.
Equation~\eqref{eq:LYZRmix} (Eq.~\eqref{eq:LYZSmix}) indicates that the intersection point of $\mathcal{R}_{nm}(T,L)$ ($L^{y_\eta}{\rm Im}\,\mu_{\rm LY}^{(n)}(T,L)$) approaches $T=T_{\rm CP}$ for $L\to\infty$. However, in this case $\mathcal{R}_{nm}(T_{\rm CP},L)$ ($L^{y_\eta}{\rm Im}\,\mu_{\rm LY}^{(n)}(T_{\rm CP},L)$) at finite $L$ has a deviation from the value for $L\to\infty$ due to the second term, and hence the intersection points between two finite $L$'s are deviated from $T=T_{\rm CP}$. Another notable consequence of Eq.~\eqref{eq:LYZmu} is that the real part of $\mu_{\rm LY}^{(n)}(T,L)$ is given by
\begin{align}
    \mathrm{Re}\, \mu_{\rm LY}^{(n)}(T,L) = \mu_{\rm CP} - \frac{a_{21}}{a_{22}} \delta T + \mathcal{O}(\delta T^2, L^{2\bar y}) \,,
    \label{eq:RemuLY}
\end{align}
which means that $\mathrm{Re}\, \mu_{\rm LY}^{(n)}(T,L)$ moves as a function of $T$ along $\vec{e}_\tau$ near the CP~\cite{Wada:2024qsk}.

To utilize the Binder cumulant in general systems, one may first define it through the baryon number cumulants \begin{align}
    \kappa_n = -\frac VT \frac{\partial^n f(T,\mu_{\rm B},L)}{\partial(\mu_{\rm B}/T)^n} \,,
    \label{eq:chi_n}
\end{align}
as~\cite{Jin:2017jjp}
\begin{align}
    \mathcal{B}_4(T,L) \equiv 
    \min_{\mu_\textrm{B}}
    \frac{\kappa_4}{\kappa_2^2}+3 \,,
    \label{eq:B4T}
\end{align}
with the free-energy density $f(T,\mu_{\rm B},L)$. The cumulants relate to the dimensionless baryon number susceptibilities via $\chi_n = \kappa_n / (VT^3)$. 

It is then follows that Eq.~\eqref{eq:B4T} behaves near the CP as 
\begin{align}
    \mathcal{B}_4(T,L) 
    &= \big( b_4 + c_4 L^{y_t} T + {\cal O}(\delta T^2) \big)
    \notag \\
    &\phantom{=.} \big( 1 + d_4 L^{\bar y} + {\cal O}(L^{2\bar y}) \big) \,.
    \label{eq:B4mix}
\end{align}
One can easily verify that Eq.~\eqref{eq:B4mix} has the same structure as Eqs.~\eqref{eq:LYZRmix} and~\eqref{eq:LYZSmix}. We, however, remark that the exponent of $L$ in Eq.~\eqref{eq:LYZRmix} on the second line is $2\bar y$, which is twice larger than those in Eqs.~\eqref{eq:LYZSmix} and~\eqref{eq:B4mix}. This indicates that the deviation of the intersection point arising from the linear mapping is suppressed more quickly for $L\to\infty$ in $\mathcal R_{nm}(T,L)$ than Eqs.~\eqref{eq:LYZSmix} and~\eqref{eq:B4mix}.
We also notice that the effect of irrelevant operators brings about further deviation of the intersection point, as discussed in Appendix~\ref{sec:irrelevant} and later sections.

\section{Model and Phase Diagram} \label{sec:model}

To investigate finite-size effects on the analytic structure of the partition function and the distribution of LYZs in the complex $\mu_{\rm B}$ plane in QCD, we employ the single-component three-color ($N_{\rm c}=3$) quark-meson model in three space dimensions ($d=3$)~\cite{Kovacs:2025gct}, whose Lagrangian density is given by
\be 
\mathcal{L}=\frac{1}{2}\partial_{\mu}\phi \partial^{\mu}\phi - U_\cl (\phi) + \bar\psi \big(i\gamma^\mu\partial_\mu- \gamma_0 \mu_q -g \phi \big)\psi \,,
\label{eq:L}
\ee
with $\psi$ and $\phi$ denoting the quark and mesonic fields, respectively, and the classical potential
\be \label{eq:Uclas}
U_\cl(\phi)=\frac{m^2}{2} \phi^2 + \frac{\lambda}{4} \phi^4-h\phi\,.
\ee
Here $\mu_q$ is the quark chemical potential, related to the baryon chemical potential by $\mu_\mathrm{B}=3\mu_q$.

Taking the mesonic field constant, $\phi(x)=\bar\phi$, and performing the integral over the fermionic field, the fermionic contribution to the effective potential in the thermodynamic limit is given by the sum of the vacuum and matter parts as 
\begin{align}
    U_{\bar qq}(\bar\phi,T,\mu_{\rm B}) &= U_{\bar qq}^\vac (\bar\phi) + U_{\bar qq}^\mat (\bar\phi,T,\mu_{\rm B}) \,,
    \label{eq:Uqq} \\
    U_{\bar qq}^\vac (\bar\phi) &= -2N_{\rm c}\int \frac{d^3 p}{(2\pi)^3} E(p)\,, 
    \label{eq:Uvac} \\
    U_{\bar qq}^\mat (\bar\phi;T,\mu_{\rm B}) &= -2N_{\rm c} T\int \frac{d^3 p}{(2\pi)^3} \left(\log g^+ + \log g^- \right)\,,
    \label{eq:Umat}
\end{align}
with $E(p)=\sqrt{p^2+g^2\bar\phi^2}$ and
\begin{align}
    g^\pm=1+e^{-(E(p)\mp \mu_q)/T}\, .
\end{align}
In the present study, we keep only the classical potential and fermionic matter part, so that the effective potential reduces to~\cite{Kovacs:2025gct} 
\be 
\mathcal{U}(\bar\phi;T,\mu_\mathrm{B})=U_\cl (\bar\phi) + U_{\bar qq}^\mat (\bar\phi;T,\mu_\mathrm{B}) \, .
\label{eq:Ueff}
\ee
It is assumed that vacuum effects are effectively encoded in the parameters of $U_{\rm cl}(\bar\phi)$. Neglecting the vacuum term removes the $-\bar\phi^4\log\bar\phi^2$ contribution, which would dominate the large-$\bar\phi$ behavior and make the potential unbounded from below. 
\begin{table}[tb]
    \centering
    \begin{tabular}{c c c | c c | c}
       & $\lambda$ & $g$ & $T_\CP$ & $\mu_{\CP}$ & $T^\mathrm{pc}_0$ \\[2pt]
        \hline\hline &&&&&\\[-.9em]
        ~set A~~ & ~29.0 ~ & ~4.55~ & ~121.53~ & ~579.97~ & ~158.46~ \\ 
        ~set B~~ & ~46.0 ~ & ~4.35~ & ~40.04~  & ~890.84~ & ~158.38~ 
    \end{tabular}
    \caption{Parameter sets used for the numerical calculations in the quark-meson model~\eqref{eq:L}. The location of the CP at $L \to\infty$, $(T,\mu_{\rm B})=(T_{\rm CP},\mu_{\rm CP})$, and the pseudocritical temperature at $\mu_\mathrm{B}=0$, defined in Eq.~\eqref{eq:mu_pcchi}, are also presented for each parameter in units of MeV.}
    \label{tab:params}
\end{table}

\begin{figure}[t]
    \centering
    \includegraphics[width=.47\textwidth]{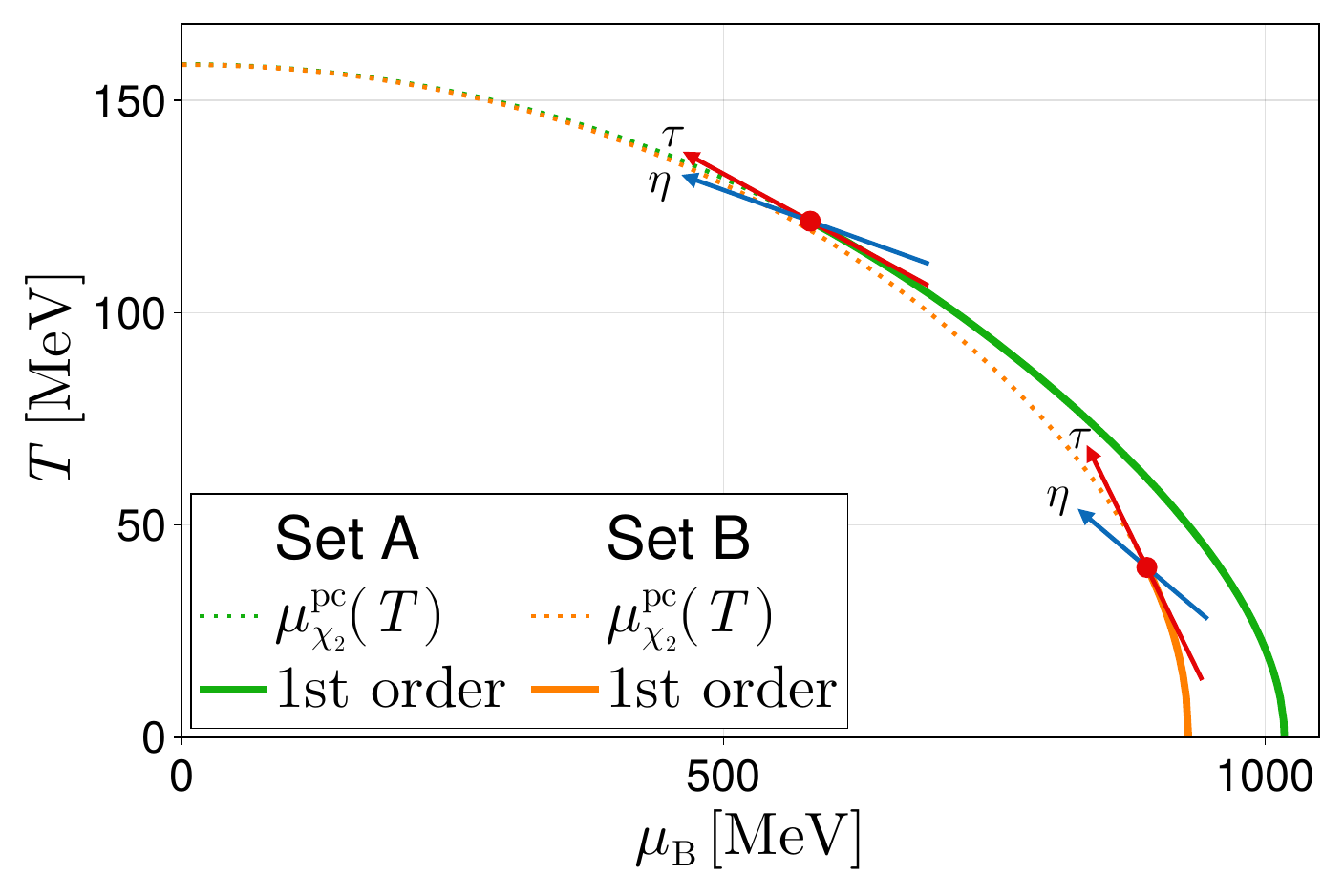}
    \caption{Phase diagram of the single-component quark-meson model for parameter sets A and B on the $T$--$\mu_\mathrm{B}$ plane. The solid and dotted lines represent the first-order phase transition and the pseudo-critical line $\mu_{\chi_2}^{\rm pc}(T)$ defined by the peak of $\chi_2$, respectively. The circle markers denote the CP. The arrows on the CP indicate the directions of the axes of Ising variables in the linear mapping~\eqref{eq:map}.}
    \label{fig:phase_diag}
\end{figure}

To determine the parameters in the model, we require that the model yields a sensible phase diagram, featuring a CP in the theoretically allowed region \cite{Philipsen:2021qji, Borsanyi:2025dyp}. We fix $m^2=-0.155~\text{GeV}^2$, $h=0.0018~\text{GeV}^3$, and choose two sets for the dimensionless parameters $\lambda$ and $g$ as listed in Tab.~\ref{tab:params}. The use of multiple parameterizations allows one to inspect the parameter dependence. 
The emerging $L\to\infty$ phase diagram for each parameter set is shown in Fig.~\ref{fig:phase_diag}. The first-order phase transition is indicated by the solid lines, while the dotted lines show the pseudo-critical line for the crossover region defined through the peak of the second-order baryon number susceptibility $\chi_2$,
\begin{align}
    \mu_{\chi_2}^{\rm pc}(T) \equiv \max_{\mu_{\rm B}} \chi_2(T,\mu_{\rm B}) \,.
    \label{eq:mu_pcchi}
\end{align}
The locations of the CP, $(T,\mu_{\rm B})=(T_{\rm CP},\mu_{\rm CP})$, denoted by the circle markers in the figure, are given in Tab.~\ref{tab:params}, together with the pseudocritical temperature $T_0^{\rm pc}$ at $\mu_{\rm B}=0$ defined by $\mu_{\chi_2}^{\rm pc}(T_0^{\rm pc})=0$. The location of the CP is numerically determined with several methods using, for example, the inflection point of $\mathcal{U}(\bar\phi)$ and the vanishing point of the real solutions of Eq.~\eqref{eq:LYES_cond}, to ensure the suppression of numerical errors. We note that while our framework with the above parameterizations provides key qualitative insights, the size scale is constrained but not completely fixed, meaning its absolute values should be interpreted with appropriate care.

As discussed in Sec.~\ref{sec:Landau}, the variables $(T,\mu_\mathrm{B})$ in the quark-meson model near the CP are related to those in the Landau potential $(\tau,\eta)$ through the linear-mapping relation, Eq.~\eqref{eq:map}. This relation can be made explicit by Taylor expanding Eq.~\eqref{eq:Ueff} with respect to $\delta\phi=\bar\phi-\bar\phi_\mathrm{CP}$, 
\begin{align}
     \mathcal U(\bar\phi, T, \mu_{\rm B})=u_0(T, \mu_{\rm B}) + \sum_{n=1}^\infty \frac{\alpha_n(T,\mu_{\rm B})}{n!} \, \delta\phi^n \,,
     \label{eq:U-exp}
\end{align}
with $\alpha_n=\partial_{\bar\phi}^n\mathcal{U}_\eff$ and $\bar\phi_{\rm CP}$ being the mean-field value of $\bar\phi$ at the CP.
At the CP, the coefficients $\alpha_n^\CP\equiv\alpha_n(T_\CP,\mu_\CP)$ vanish for $n=1,2,3$, while they are nonzero for $n>3$ in general. In the vicinity of the CP, therefore, Eq.~\eqref{eq:U-exp} is further expanded as 
\begin{align}
    &\mathcal U(\bar\phi, T, \mu_\mathrm{B})\notag\\
    &=
    u_0- (a_{21}\delta T + a_{22}\delta \mu_\mathrm{B}) \delta\bar\phi + \frac{a_{11}\delta T + a_{12}\delta \mu_\mathrm{B}}{2} \, \delta\bar\phi^2
    \notag \\
    & \phantom{=.} + \frac{b}{4!}\,\delta\bar\phi^4 + \sum_{n=5}^\infty \frac{\alpha_n^{\rm CP}}{n!}\,\delta \bar\phi^n \,,
    \label{eq:U-exp2}
\end{align}
at the leading order in $\delta T$ and $\delta\mu_{\rm B}$, where the third-order term is eliminated by a shift of $\bar\phi_{\rm CP}$ without altering the integral measure of Eq.~\eqref{eq:ZMF}. The comparison of Eqs.~\eqref{eq:Landau}, \eqref{eq:map} and~\eqref{eq:U-exp2} shows that the coefficients $a_{ij}$ and $b$ in Eq.~\eqref{eq:U-exp2} are identical to those in Eq.~\eqref{eq:map}. Consequently, these coefficients are given by the mixed partial derivatives of $\mathcal{U}$ evaluated at the CP
\begin{align}
 a_{11}=\partial_{T}\partial_{\bar\phi}^2\, \mathcal{U}\,,& \quad  a_{12}=\partial_{\mu_\mathrm{B}}\partial_{\bar\phi}^2\, \mathcal{U}\,, \notag \\ a_{21}=-\partial_{T}\partial_{\bar\phi}\, \mathcal{U} \,,& \quad a_{22}=-\partial_{\mu_\mathrm{B}}\partial_{\bar\phi}\, \mathcal{U}\,, \label{eq:map_elemets}\\
  b =& \partial_{\bar\phi}^4\, \mathcal U\,,
  \notag
\end{align}
in line with Refs.~\cite{Pradeep:2019ccv, Parotto:2018pwx}. 
In Fig.~\ref{fig:phase_diag}, the directions of the $\tau$ and $\eta$ axes in the linear mapping, Eq.~\eqref{eq:axes}, obtained through Eq.~\eqref{eq:map_elemets} are depicted on the CP for each parameter set.

In Eq.~\eqref{eq:U-exp2}, higher-order terms $\alpha_n^{\rm CP}$ are generally nonvanishing for $n\ge4$ even for odd orders. This is in contrast to the case where the $Z_2$ symmetry of the invariance under $\bar\phi\to-\bar\phi$ is imposed in Eq.~\eqref{eq:Landau}. We will see that the existence of the $\delta\bar\phi^5$ term gives rise to a nontrivial exponent in the scaling relations in later sections.

For finite $L$, the momentum integral in Eq.~\eqref{eq:Uqq} is replaced with the sum over the discretized momenta. As already discussed in Sec.~\ref{sec:Landau}, this effect brings about the $L$ dependence of ${\cal U}(\bar\phi,T,\mu_{\rm B})$. However, the resulting $L$ dependence of the observables is non-analytic as discussed in Sec.~\ref{sec:Landau}, and its effect in Eq.~\eqref{eq:Umat} is exponentially suppressed as $e^{-c/L}$ for increasing $L$. 
In fact, the momentum space constraints on the fermionic fluctuations have a significant effect only for small $L$~\cite{Palhares:2009tf, Kovacs:2023kbv}, which lie at the lower end of our region of interest in most applications. Therefore, in the following we use the $L\to\infty$ effective potential, with finite-size effects entering only through its prefactor in Eq.~\eqref{eq:ZMF}. We, however, note that the implicit $L$ dependence of the potential is important for yet small $L$, and in this region the effects can strongly depend on the chosen boundary conditions, treatment of ${\cal U}_{\rm vac}(\bar\phi)$ in Eq.~\eqref{eq:Uqq}, and the approximation used~\cite{Kovacs:2023kbv}. 

\section{Lee-Yang zeros and edge singularity} \label{sec:LYZandES}

In this section, we discuss the behavior of the LYZs and LYES in our model. For the numerical determination of LYZs, we introduce the normalized partition function
\begin{align}
    \mathcal{Z}_{\rm N}(T,\mu_\textrm{B},L) \equiv \frac{\mathcal{Z}(T,\mu_\textrm{B},L)}{\mathcal{Z}(T,\mathrm{Re}\,\mu_\textrm{B},L)} \,,
    \label{eq:normalized_Z}
\end{align}
and solve the simultaneous equations
\begin{align}
    {\rm Re}\, \mathcal{Z}_{\rm N}(T,\mu_\textrm{B},L)=0 \,,
    &&
    {\rm Im}\, \mathcal{Z}_{\rm N}(T,\mu_\textrm{B},L)=0 \,,
    \label{eq:ReIm}
\end{align}
for complex $\mu_\textrm{B}$ with fixed $T$ and $L$. Whereas the normalization in Eq.~\eqref{eq:normalized_Z} does not alter the final result since the denominator is real and positive, it acts to improve numerical stability. To solve Eq.~\eqref{eq:ReIm} numerically, we set an initial guess for each LYZ and search for the solution around it iteratively. For determining the initial values, we make the contour map of $|\mathcal{Z}_{\rm N}(T,\mu_\textrm{B},L)|$ in the complex $\mu_\textrm{B}$ plane and find approximate values of individual LYZs satisfying $|\mathcal{Z}_{\rm N}(T,\mu_\textrm{B},L)|=0$. For a search for LYZs at a continuous range of $T$, we vary $T$ with a small interval $\Delta T$ and use the LYZs obtained at the previous $T$ value for the initial guess of the next step. We also found that this procedure can be further improved by employing the initial guess obtained through a linear extrapolation using two previous $T$ values.
The LYES are determined by solving Eq.~\eqref{eq:LYES_cond} for $T>T_{\rm CP}$ for complex $\bar\phi$ and $\mu_{\rm B}$. 

\begin{figure}[t]
    \centering
    \includegraphics[width=.47\textwidth]{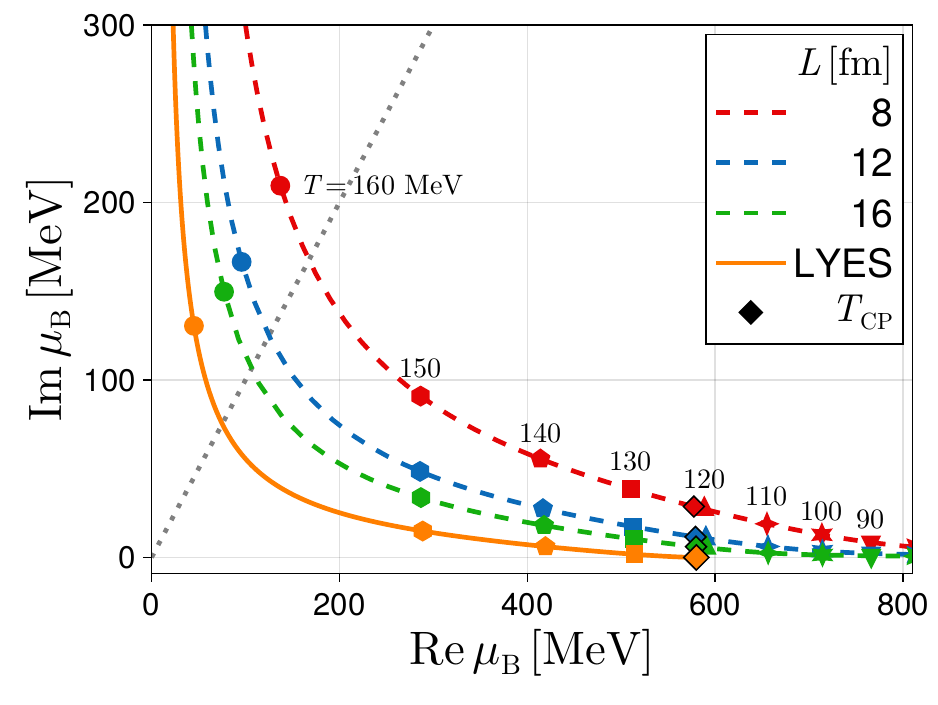}
    \includegraphics[width=.47\textwidth]{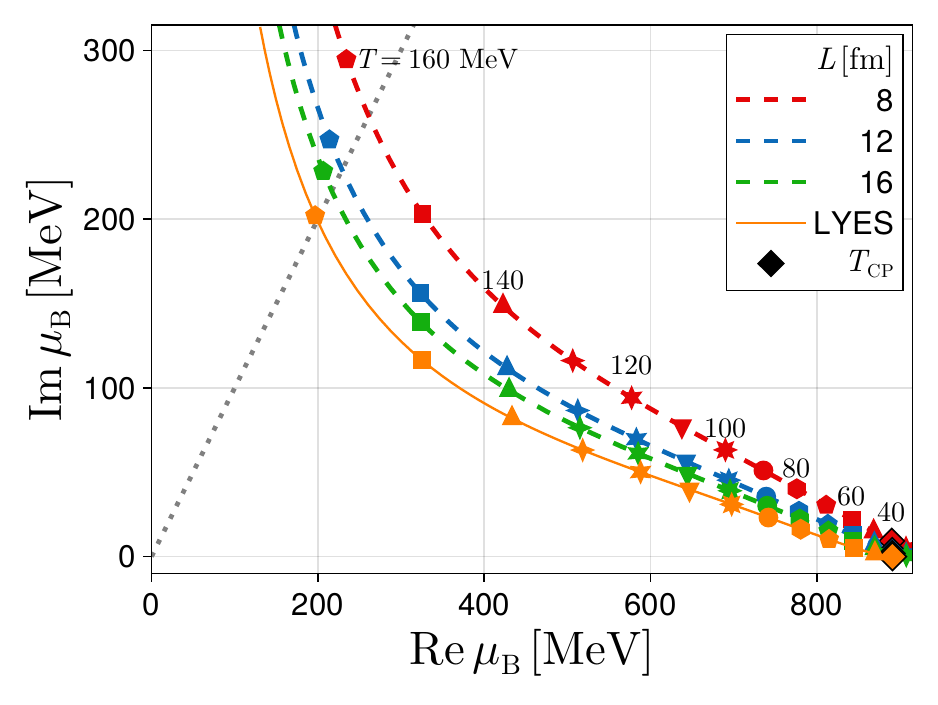}
    \caption{
    Trajectories of the first LYZs $\mu_{\rm LY}^{(1)}(T,L)$ for $L = 8,\,12,\,16~\mathrm{fm}$ (dashed lines) and the LYES $\mu_{\rm ES}(T)$ (solid line) in the complex $\mu_{\rm B}$ plane for the parameter set~A (top) and set~B (bottom). The locations at several temperatures are indicated by different markers. The dotted line represents $\mathrm{Im}\,\mu_{\rm B} = \mathrm{Re}\,\mu_{\rm B}$.
    }
    \label{fig:LYES_LYZ_L}
\end{figure}

    In Fig.~\ref{fig:LYES_LYZ_L}, we show the trajectories of the first LYZ, $\mu_{\rm LY}^{(1)}(T,L)$, for various $L$ and the LYES, $\mu_{\rm ES}(T)$, with the variation of $T$ in the complex $\mu_{\rm B}$ plane. The upper and lower panels show the results for the parameter sets~A and~B, respectively. The dashed lines show the trajectories of the first LYZ at $L=8,\,12,\,16~\mathrm{fm}$, while the solid line represents the LYES. The markers on the lines denote their locations at various temperatures; the diamond markers indicate the locations at $T=T_{\textrm{CP}}$. 
The figure shows that the LYZ converges to the LYES (real axis) for $T>T_{\textrm{CP}}$ ($T<T_{\textrm{CP}}$) in the $ L \rightarrow \infty$ limit, as discussed in Sec.~\ref{sec:LYZ}.
One also finds that the real part of $\mu_{\rm LY}^{(1)}(T,L)$ is insensitive to $L$ near the CP. This behavior indicates that the $\mathcal{O}(L^{2\bar y})$ term in Eq.~\eqref{eq:RemuLY} is well suppressed.

\begin{figure}[t]
    \centering
    \includegraphics[width=.47\textwidth]{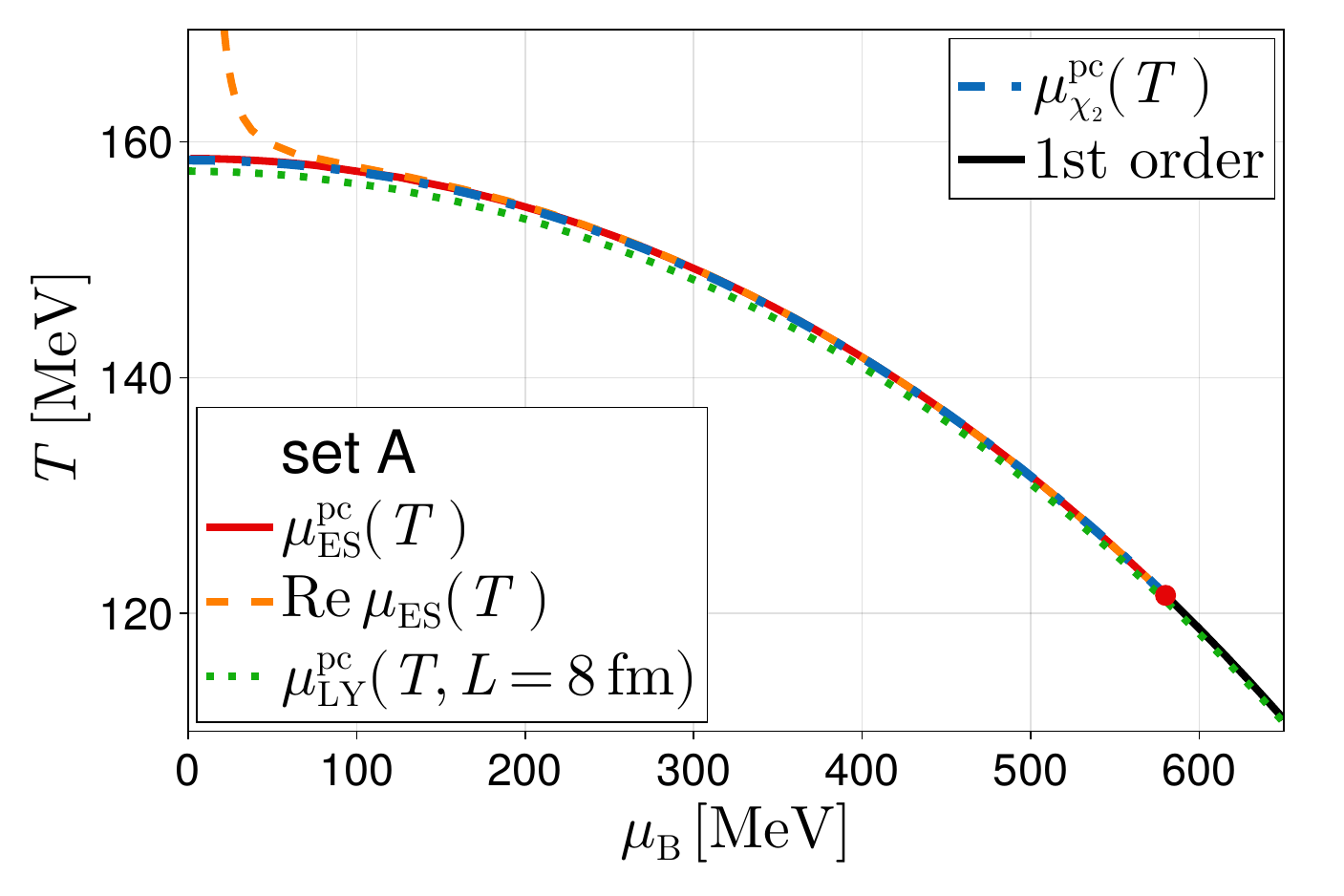}
    \caption{Comparison of various definitions of the pseudo-critical line, $\mu_{\chi_2}^{\rm pc}(T)$, $\mu_{\rm ES}^{\rm pc}(T)$, $\mu_{\rm LY}^{\rm pc}(T,L)$ at $L=8$~fm, and ${\rm Re}\mu_{\rm ES}(T)$, for parameter set A. The black solid line represents the first-order phase transition.}
    \label{fig:phase_diag2}
\end{figure}

With increasing $T$, the trajectory of $\mu_{\rm ES}(T)$ curves toward an upward direction for $T\gtrsim155$~MeV near the imaginary axis in both results. The first LYZ $\mu_{\rm LY}^{(1)}(T,L)$ also follows this behavior, and ${\rm Re}\,\mu_{\rm LY}^{(1)}(T,L)$ starts to have a strong $L$ dependence in this range of $T$. 
In Fig.~\ref{fig:phase_diag2}, we plot the trajectory of ${\rm Re}\,\mu_{\rm ES}(T)$ on the $T$--$\mu_{\rm B}$ plane by the dashed line together with $\mu_{\chi_2}^{\rm pc}(T)$ in Eq.~\eqref{eq:mu_pcchi} shown by the dash-dotted line for the parameter set~A. The figure shows that they agree well for $T\lesssim155$~MeV, while ${\rm Re}\,\mu_{\rm ES}(T)$ suddenly deviates from $\mu_{\chi_2}^{\rm pc}(T)$ and jumps up near the $T$-axis. A similar result is also observed for the parameter set~B. These behaviors of the LYES qualitatively agree with those reported in previous studies~\cite{Mukherjee:2021tyg, Zhang:2025jyv, Wan:2025wdg}. 

\begin{figure}[t]
    \centering
    \includegraphics[width=.47\textwidth]{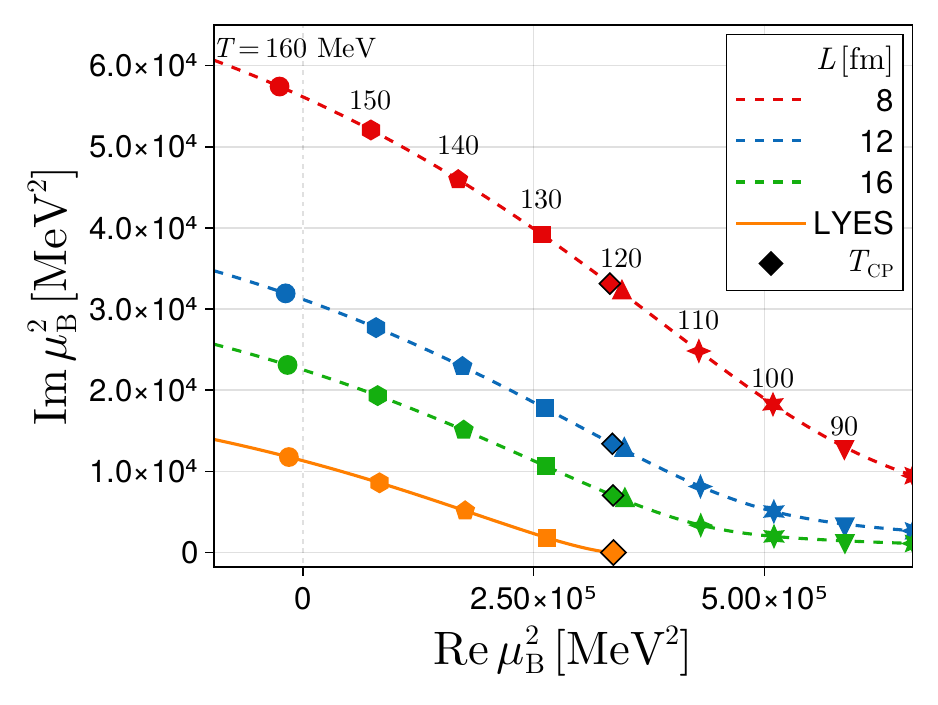}
    \caption{Trajectories of the first LYZs $\mu_\textrm{LY}^{(1)}(T,L)$ for $L = 8,\,12,\,16~\mathrm{fm}$ (dashed lines) and the LYES $\mu_\textrm{ES}(T)$ (solid line) in the complex $\mu_\mathrm{B}^2$ plane. The meanings of lines and markers are the same as those in Fig.~\ref{fig:LYES_LYZ_L}.}
    \label{fig:LYES_LYZ_squared}
\end{figure}

These results on the LYZs and LYES at small ${\rm Re}\,\mu_{\rm B}$ are nicely understood as follows. First, because of the charge conjugation symmetry, our partition function is an even function satisfying ${\cal Z}(T,\mu_{\rm B},L)={\cal Z}(T,-\mu_{\rm B},L)$, and it depends on $\mu_{\rm B}$ only through $\mu_{\rm B}^2$. To respect this property, the linear mapping relation in Eq.~\eqref{eq:map} may be replaced by
\begin{align}
    \begin{pmatrix} \tau \\ \eta \end{pmatrix} = 
    \begin{pmatrix} a_{11} & a_{12}/2\mu_{\rm CP} \\ a_{21} & a_{22}/2\mu_{\rm CP}\end{pmatrix}
    \begin{pmatrix} T-T_{\rm CP} \\ \mu_{\rm B}^2 -\mu_{\rm CP}^2 \end{pmatrix} .
    \label{eq:map2}    
\end{align}
Although Eq.~\eqref{eq:map2} is equivalent with Eq.~\eqref{eq:map} in the vicinity of the CP, Eq.~\eqref{eq:map2} would provide a better description for a wider range of $T$ and $\mu_{\rm B}$. With a replacement of Eq.~\eqref{eq:map} with Eq.~\eqref{eq:map2}, Eq.~\eqref{eq:RemuLY} is modified as
\begin{align}
    \mathrm{Re} \big(\mu_{\rm LY}^{(n)}(T,L)\big)^2 = \mu_{\rm CP}^2 - 2\mu_{\rm CP} \frac{a_{21}}{a_{22}} \delta T + \mathcal{O}(\delta T^2, L^{2\bar y}) \, .
    \label{eq:RemuLY2}
\end{align}
Equation~\eqref{eq:mu_ES} is also modified accordingly.

To see how Eq.~\eqref{eq:map2} works, in Fig.~\ref{fig:LYES_LYZ_squared} we plot $\mu_{\rm LY}^{(1)}(T,L)$ and $\mu_{\rm ES}(T)$ in the complex $\mu_{\rm B}^2$ plane for the parameter set~A. The meanings of the lines and markers are the same as those in Fig.~\ref{fig:LYES_LYZ_L}. 
From the figure, one sees that the LYZs and LYES behave more linearly for $T\gtrsim155$~MeV than Fig.~\ref{fig:LYES_LYZ_L}. Moreover, ${\rm Re}\big(\mu_{\rm LY}^{(1)}(T,L)\big)^2$ is insensitive to $L$ even for $T\gtrsim155$~MeV, which is interpreted as the suppression of ${\cal O}(L^{2\bar y})$ term in Eq.~\eqref{eq:RemuLY2}. 
We notice that the difference between Eqs.~\eqref{eq:map} and~\eqref{eq:map2} becomes prominent at small ${\rm Re}\,\mu_{\rm B}$. For ${\rm Re}\mu_{\rm B}\gg {\rm Im}\mu_{\rm B}\ge0$, one has ${\rm Re}\mu_{\rm B}^2\simeq({\rm Re}\mu_{\rm B})^2$ and ${\rm Im}\mu_{\rm B}^2\simeq 2{\rm Re}\mu_{\rm B}{\rm Im}\mu_{\rm B}$, which make the mapping between the complex $\mu_{\rm B}$ and $\mu_{\rm B}^2$ planes simple. However, these relations are violated when ${\rm Im}\,\mu_{\rm B}$ becomes comparable with ${\rm Re}\,\mu_{\rm B}$. For the LYZs and LYES in our model, this occurs for $T\gtrsim155$~MeV. In Fig.~\ref{fig:LYES_LYZ_L}, the line of ${\rm Re}\,\mu_{\rm B} = {\rm Im}\,\mu_{\rm B}$, i.e., ${\rm Re}\,\mu_{\rm B}^2=0$, is shown by the dotted line. One sees that the trajectories exhibit clear upward bending around this line. 

The deviation of ${\rm Re}\,\mu_{\rm ES}(T)$ from the susceptibility peak at small ${\rm Re}\,\mu_{\rm B}$ in Fig.~\ref{fig:phase_diag2} can also be understood in a similar manner.
Equation~\eqref{eq:map2} also suggests that the pseudo-critical line corresponding to ${\rm Re}\,\eta=0$ defined throug $\mu_{\rm B}^2$ as
\begin{align}
    \mu_{\rm ES}^{\rm pc}(T)\equiv \sqrt{{\rm Re}\,\mu_{\rm ES}(T)^2} \,,
    \label{eq:mu_pcES}
\end{align}
has a better agreement with $\mu_{\chi_2}^{\rm pc}(T)$.
In Fig.~\ref{fig:phase_diag2}, the trajectory of $\mu_{\rm ES}^{\rm pc}(T)$ is shown by the solid line, which agrees quite well with $\mu_{\chi_2}^{\rm pc}(T)$~\cite{Ejiri:2014oka}, suggesting that Eq.~\eqref{eq:map2} indeed describes the mapping well for $\mu_{\rm B}\simeq0$.

The fact that ${\rm Re}\big(\mu_{\rm LY}^{(1)}(T,L)\big)^2$ is insensitive to $L$, as in Fig.~\ref{fig:LYES_LYZ_squared}, also indicates that 
\begin{align}
    \mu_{\rm LY}^{\rm pc}(T,L)\equiv \sqrt{{\rm Re}\big(\mu_{\rm LY}^{(1)}(T,L)\big)^2} \,,
    \label{eq:mu_pcLY}
\end{align}
at finite $L$ can be an alternative definition of the pseudo-critical line. As shown in Fig.~\ref{fig:phase_diag2}, Eq.~\eqref{eq:mu_pcLY} indeed agrees well with $\mu_{\chi_2}^{\rm pc}(T)$ and $\mu_{\rm ES}^{\rm pc}(T)$ even at $L=8$~fm. To take the $L\to\infty$ limit of $\mu_{\rm LY}^{\rm pc}(T,L)$,
the fact that its leading $L$ dependence is given as in Eq.~\eqref{eq:RemuLY2} would play a useful role.
A similar result is also obtained for the parameter set~B, whereas the deviation of the different definitions of the pseudo-critical lines becomes slightly larger around $\mu_{\rm B}\simeq0$.

\begin{figure}[t]
    \centering
    \includegraphics[width=.47\textwidth]{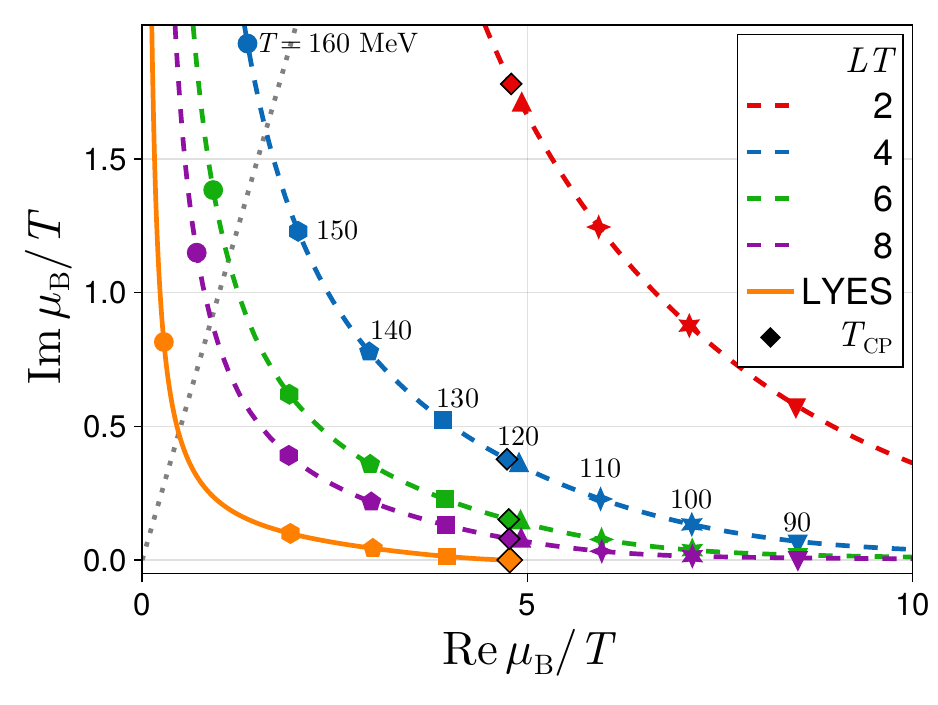}
    \caption{
    Trajectories of the first LYZs for $LT = 2,\,4,\,6,\,8~\mathrm{fm}$ (dashed lines) and the LYES (solid line) in the complex $\mu_\mathrm{B}/T$ plane.
    The dotted line corresponds to $\mathrm{Im}\,\mu_\mathrm{B}/T = \mathrm{Re}\,\mu_\mathrm{B}/T$.}
    \label{fig:LYES_LYZ_LT}
\end{figure}

It is instructive to make a comparison of our results on the LYZ with those obtained in lattice-QCD numerical simulations via Padé approximation~\cite{Bollweg:2022rps, Basar:2023nkp, Clarke:2024ugt, Adam:2025phc}. For this purpose, in Fig.~\ref{fig:LYES_LYZ_LT} we plot the trajectories of the first LYZ with fixed aspect ratios $LT=2,4,6,8$, together with the LYES in the complex $\mu_{\rm B}/T$ plane for the parameter set~A. From the figure, one sees that the deviation of the first LYZ from the LYES is clear even at $LT=8$ and grows large with decreasing $LT$. Although the first LYZ obtained on the lattice is sometimes regarded as the LYES in the literature, Fig.~\ref{fig:LYES_LYZ_LT} suggests that such a substitution has a large error at the $LT$ employed in these simulations. An important qualitative difference between the LYES and LYZ found in Fig.~\ref{fig:LYES_LYZ_LT} is that ${\rm Im}\,\mu_{\rm ES}(T)$ vanishes at $T=T_{\rm CP}$, while ${\rm Im}\,\mu_{\rm LY}^{(n)}(T,L)$ never touches the real axis. 
In Refs.~\cite{Basar:2023nkp, Clarke:2024ugt, Adam:2025phc}, the lattice results of $\mu_{\rm LY}^{(1)}(T,L)$ obtained at several temperatures for $T>T_{\rm CP}$ are extrapolated to lower $T$, and the value of $T$ satisfying ${\rm Im}\,\mu_{\rm LY}^{(1)}(T,L)=0$ is identified as the $T_{\rm CP}$. Figure~\ref{fig:LYES_LYZ_LT} suggests that this analysis underestimates $T_{\rm CP}$ since ${\rm Im}\,\mu_{\rm LY}^{(1)}(T,L)$ is always larger than ${\rm Im}\,\mu_{\rm ES}(T)$; in other words, the value of $\mu_{\rm CP}$ is expected to be overestimated in this analysis. In any case, FSS of the LYZ plays a crucial role in properly identifying $T_{\rm CP}$ from $\mu_{\rm LY}^{(n)}(T,L)$~\cite{Wada:2024qsk,Wada:2025ycz}. We will address this issue in the next section.

The lattice results in Refs.~\cite{Basar:2023nkp, Clarke:2024ugt, Adam:2025phc} also show that the real part of the first LYZ tends to stay at a large value ${\rm Re}\,\mu_{\rm LY}^{(1)}(T,L)/T\simeq3$ even near $T^{\rm pc}_{\mu=0}$ (See, Fig.~3 in Ref.~\cite{Clarke:2024ugt}, for example). Comparing this behavior with Fig.~\ref{fig:LYES_LYZ_LT}, it is suggested that they are to a large extent attributed to the finite-size effects, although a possible non-monotonic $T$ dependence of ${\rm Re}\,\mu_{\rm LY}^{(1)}(T,L)/T$ found in these results is not observed in our result.
We note that the absence of Roberge-Weiss (RW) symmetry of QCD~\cite{Roberge:1986mm} in our model might be the origin of the difference, while the analysis in Ref.~\cite{Wan:2025wdg} suggests that its effect on the LYES is not significant.

\section{Intersection analysis} \label{sec:CP}

In Sec.~\ref{sec:intersection}, we introduced several methods to locate the CP through the intersection analysis using the numerical results at finite $L$. 
In this section, we test their validity in our model. Throughout this section, we present the results for the parameter set~A, as the numerical results hardly change qualitatively with the variation of parameters. 

\begin{figure}[t]
    \centering
    \includegraphics[width=.47\textwidth]{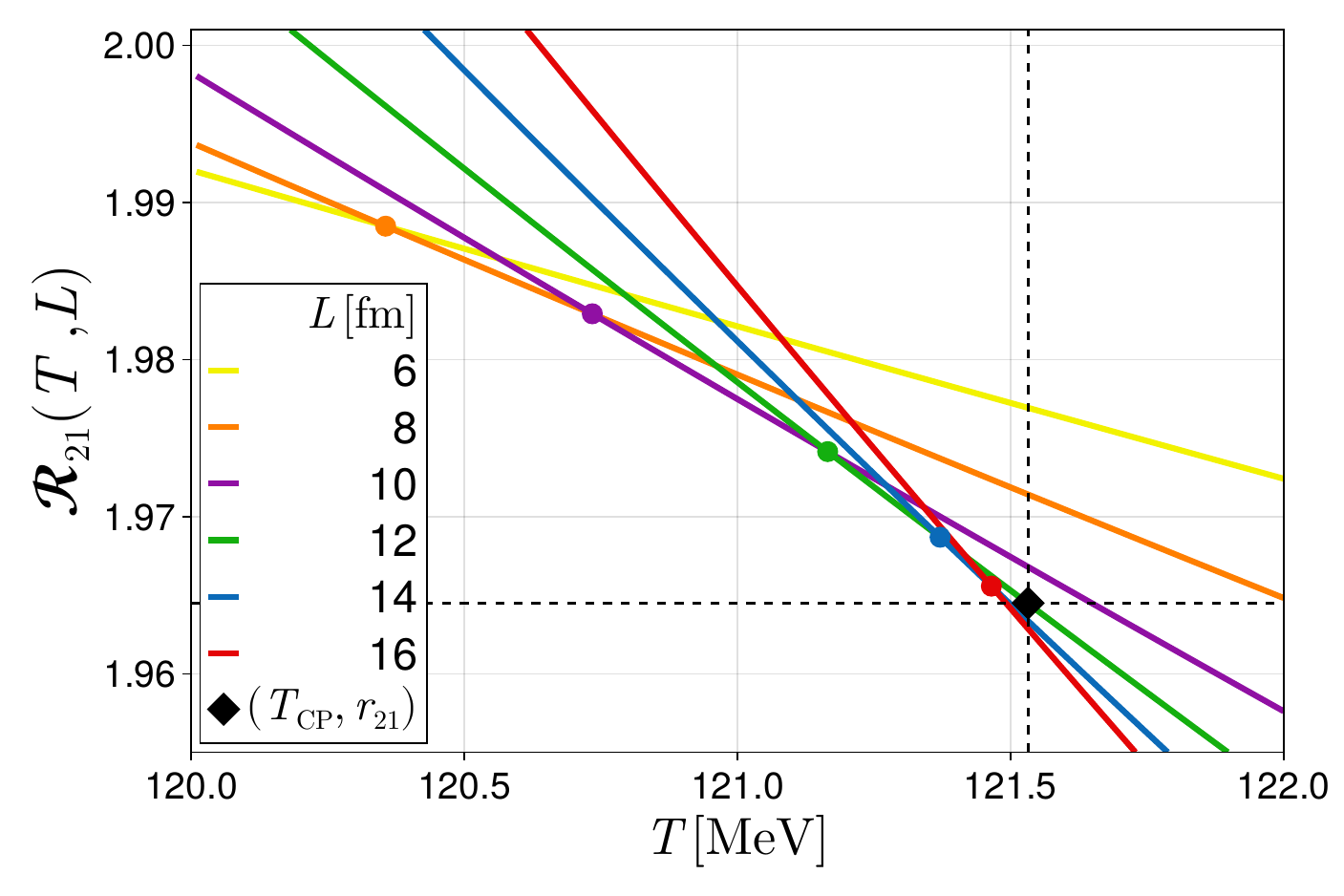}
    \includegraphics[width=.47\textwidth]{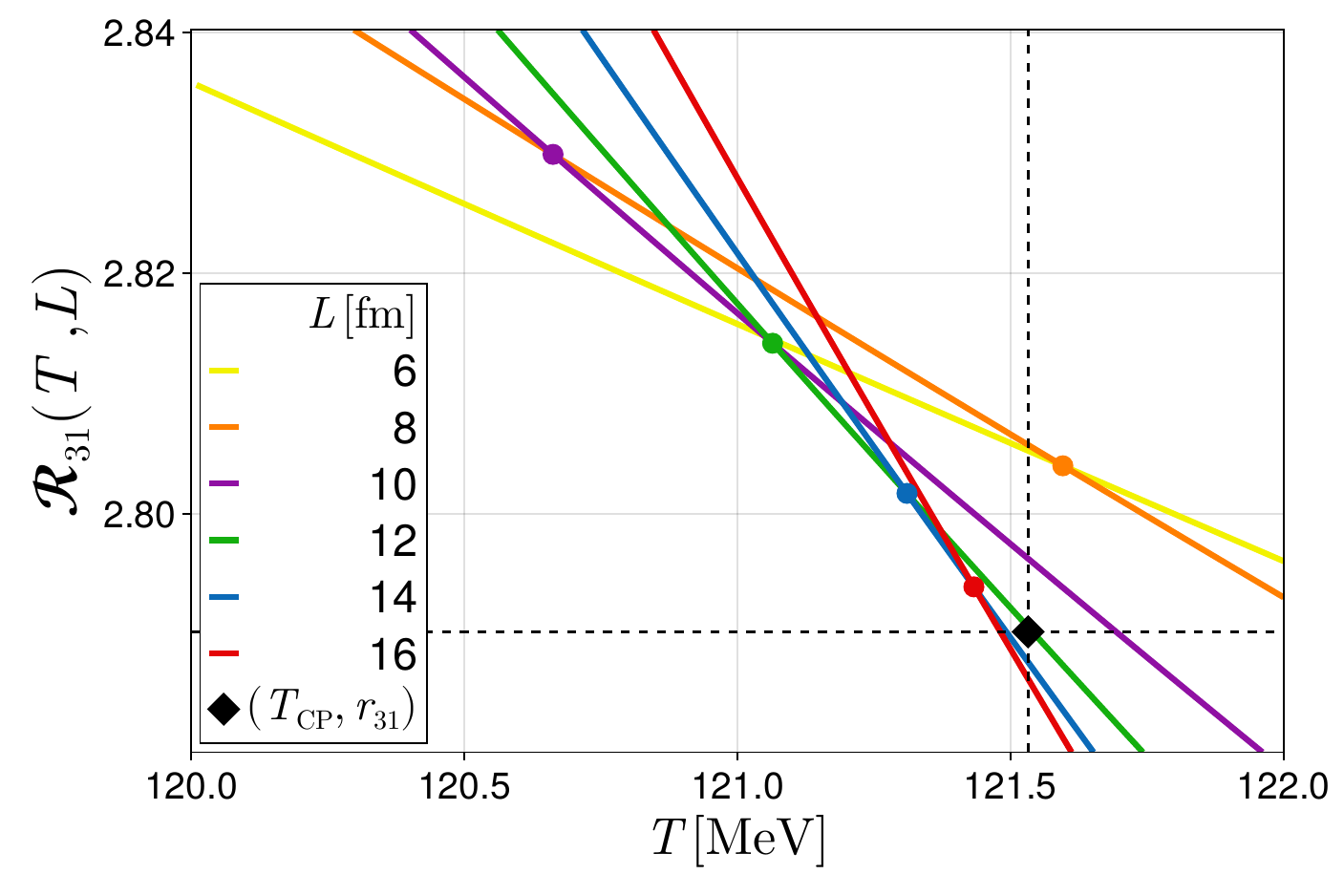}
    \includegraphics[width=.47\textwidth]{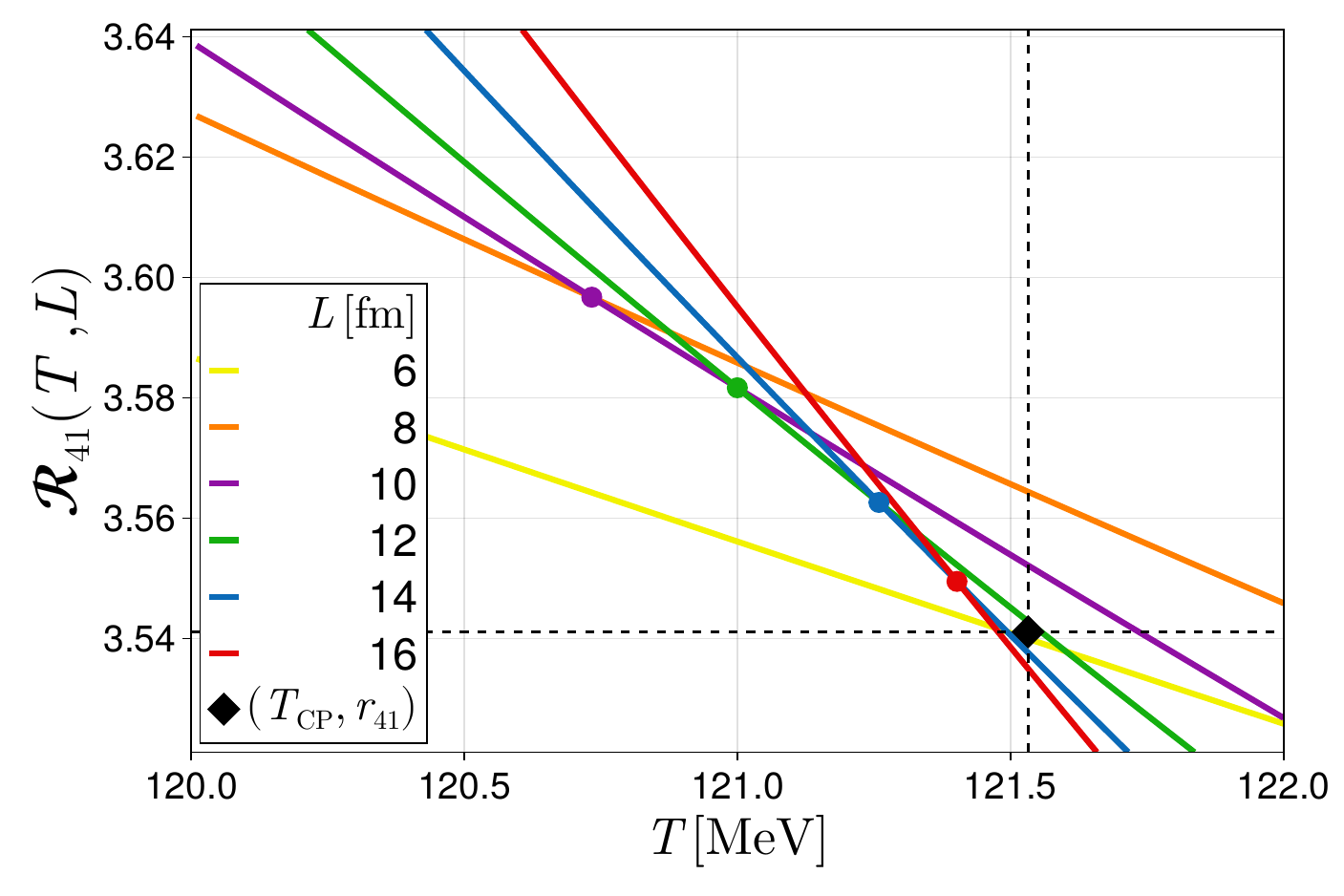}
    \caption{Lee--Yang zero ratios $\mathcal{R}_{n1}(T,L)$ at $L = 6, 8, 10, 12, 14,$ and $16~\mathrm{fm}$. From top to bottom, the panels show $\mathcal{R}_{21}(T,L)$, $\mathcal{R}_{31}(T,L)$, and $\mathcal{R}_{41}(T,L)$, respectively. Markers indicate intersection points for pairs of adjacent volumes, with their colors corresponding to the larger volume in each pair. The vertical and horizontal dashed lines indicate $T_\mathrm{CP}$ and the value of the intersection point $r_{n1}$ for $L\to\infty$, respectively.}
    \label{fig:LYZR}
\end{figure}

In Fig.~\ref{fig:LYZR}, we show the behavior of the LYZRs ${\cal R}_{n1}(T,L)$ near $T=T_{\rm CP}$ at $L = 6,\,8,\,\ldots,16~\mathrm{fm}$ for $n=2,3,4$ from the top to bottom panels, respectively. The intersection points between two neighboring $L$'s are indicated by the circle markers; the color of each marker corresponds to that of the larger $L$. In the figure, $T=T_{\rm CP}$ and the mean-field values of $r_{n1}$, Eq.~\eqref{eq:r_n1_MF}, are indicated by the vertical and horizontal dashed lines, respectively, with the diamond symbol denoting their crossing point. One finds that the intersection point of ${\cal R}_{n1}(T,L)$ for different $L$ moves toward the diamond symbol with increasing $L$, as discussed in Sec.~\ref{sec:intersection}. The values of $T$ at the intersection points are within $1\%$ of $T_{\rm CP}$ when the LYZR for $L\ge8$~fm ($LT\gtrsim5$) are used for all $n$, and the precision improves with increasing $L$. On the other hand, ${\cal R}_{n1}(T,L)$ at $L=6$~fm, shown by the yellow line, clearly deviates from the trend of the other lines. This behavior, which is more pronounced for larger $n$, may come from the finite-size effects beyond FSS, Eqs.~\eqref{eq:Zscaling} and~\eqref{eq:fscaling}.

\begin{figure}[t]
    \centering
    \includegraphics[width=.47\textwidth]{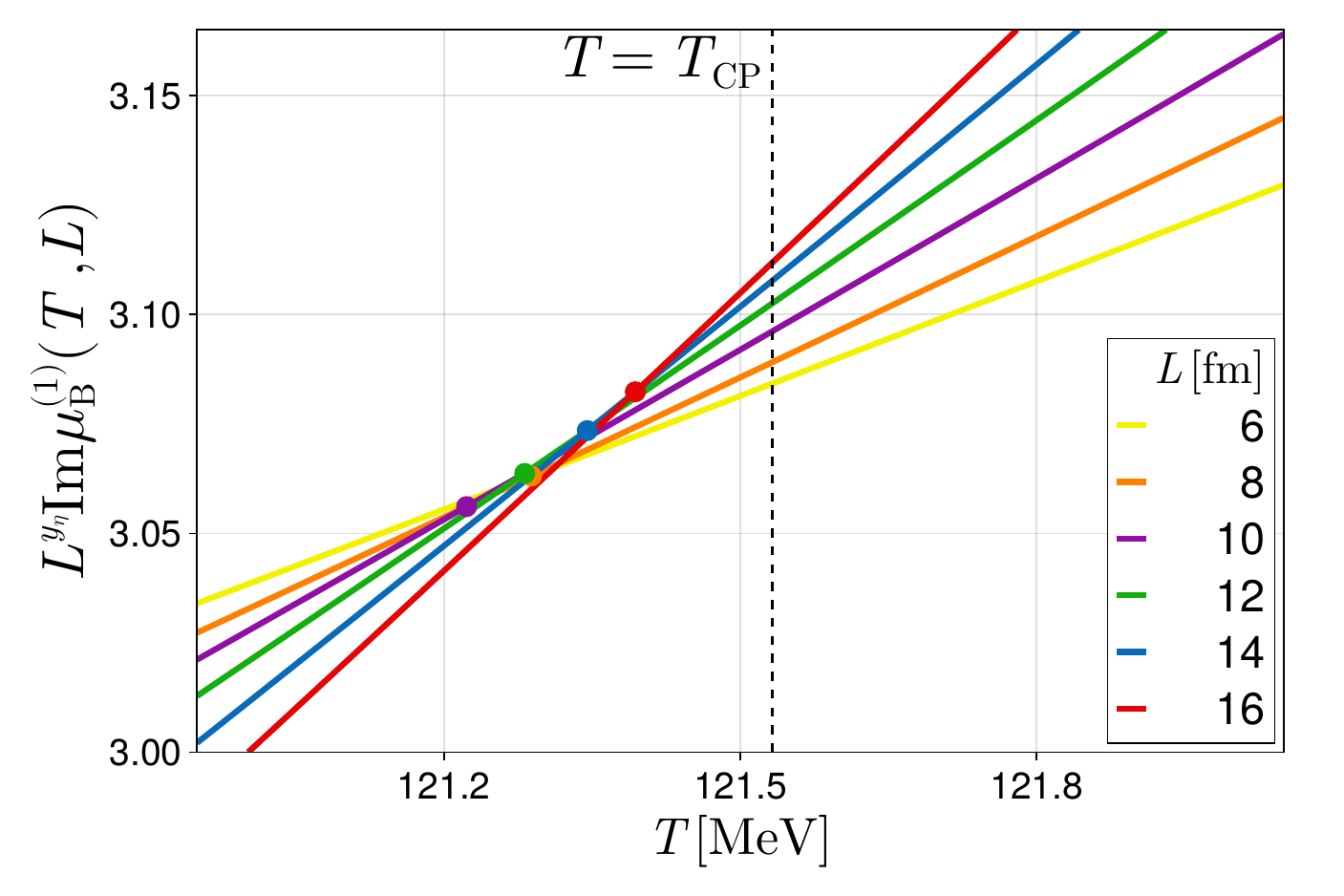}
    \includegraphics[width=.47\textwidth]{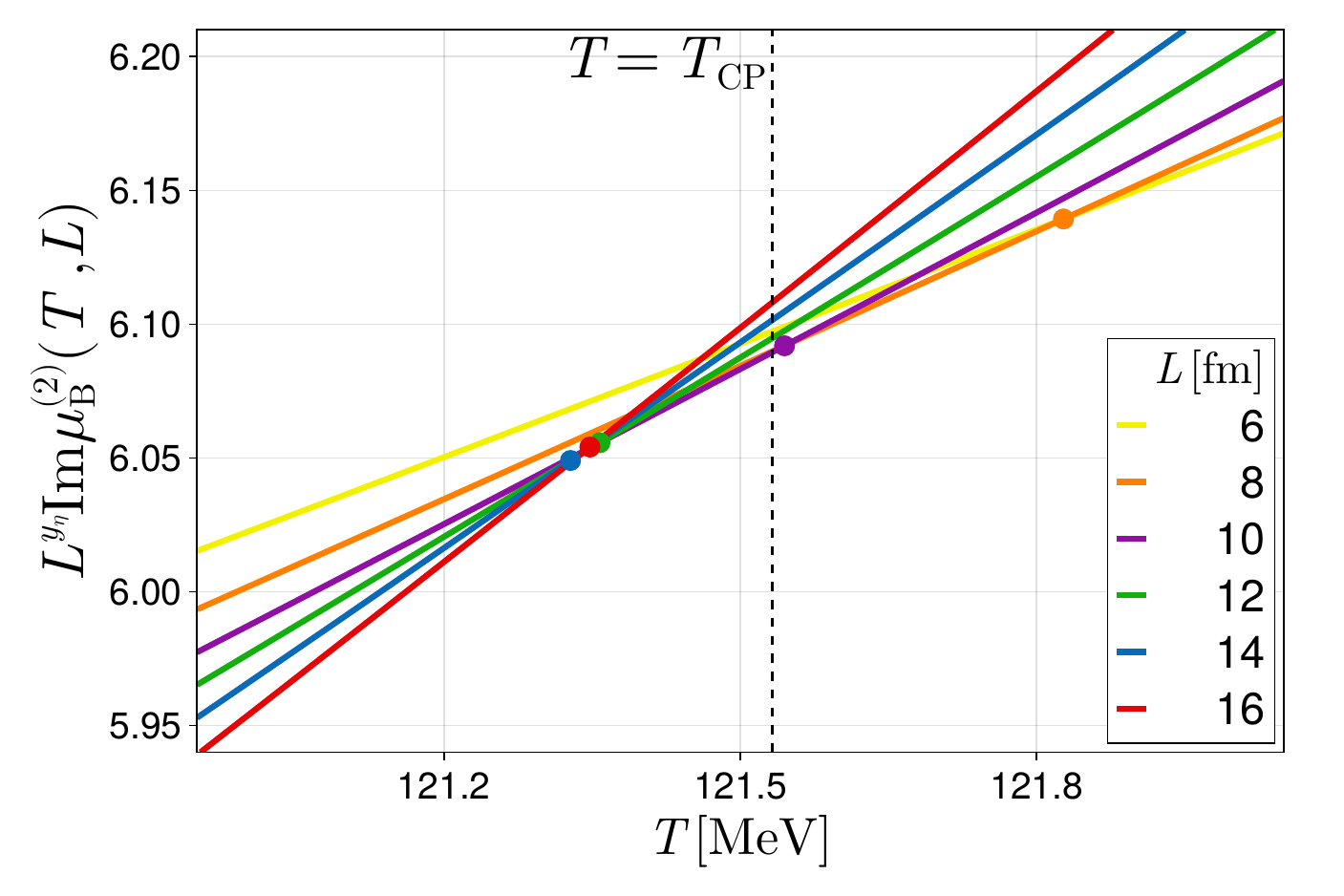}
    \caption{Scaled LYZs $L^{y_\eta}\textrm{Im}\,\mu_\mathrm{LY}^{(n)}(T,L)$ at $L = 6, 8, 10, 12, 14,$ and $16~\mathrm{fm}$, with the top and bottom panels corresponding to $n = 1$ and $n = 2$, respectively.}
    \label{fig:LYZscales}
\end{figure}

\begin{figure}[t]
    \centering
    \includegraphics[width=.47\textwidth]{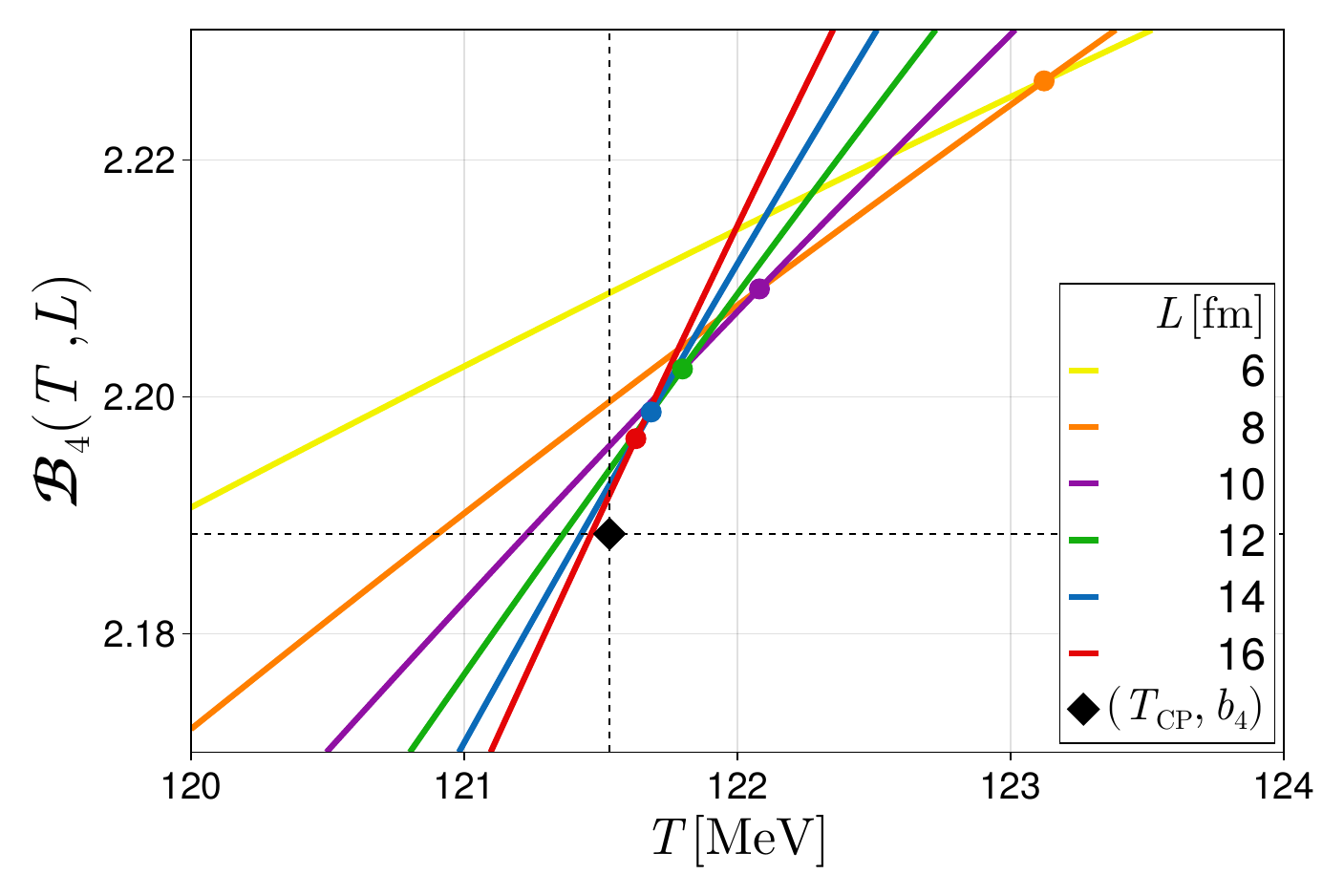}
    \caption{Fourth-order Binder cumulant $\mathcal{B}_4(T,L)$ for $L = 6, 8, \cdots,16~\mathrm{fm}$. 
    The horizontal dashed line indicates $b_4$, the value of the intersection point for $L\to\infty$.}
    \label{fig:B4}
\end{figure}

In Figs.~\ref{fig:LYZscales} and~\ref{fig:B4}, we show the same plots for the scaled LYZ $L^{y_\eta}\textrm{Im}\,\mu^{(n)}_\textrm{LY}(T,L)$ for $n=1,2$ (top and bottom panels) and the Binder cumulant ${\cal B}_4(T,L)$ defined, respectively, in Eqs.~\eqref{eq:LYZSmix} and~\eqref{eq:B4T}. For the scaled LYZ, we use the mean-field critical exponent $y_\eta$ in Eq.~\eqref{eq:y}. The limiting value of the intersection point of ${\cal B}_4(T,L)$ in the mean-field models, Eq.~\eqref{eq:b4_MF}, is indicated by the horizontal dashed line in Fig.~\ref{fig:B4}. Since the limiting value for $L^{y_\eta}\textrm{Im}\,\mu^{(n)}_\textrm{LY}(T,L)$ is not universal, Fig.~\ref{fig:LYZscales} does not have the corresponding line. The figures show that the intersection points moves towards $T_{\rm CP}$ in these quantities similarly to the LYZRs, suggesting that all the intersection analyses are equally powerful to locate the CP, although the convergence is slightly faster (slower) in the scaled LYZ (Binder cumulant) in our model. 

\begin{figure}[t]
    \centering
    \includegraphics[width=.47\textwidth]{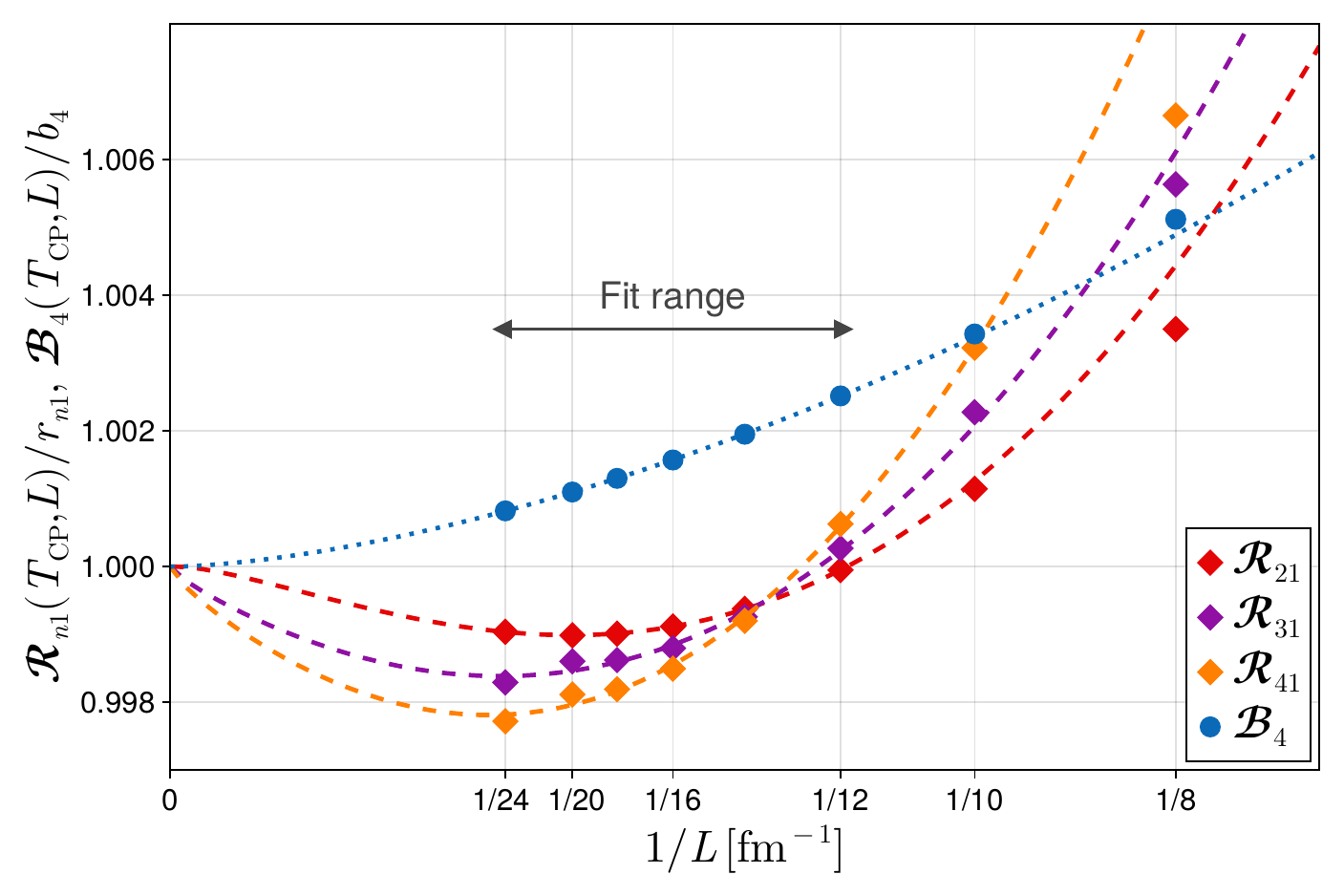}
    \caption{System-size dependence of $\mathcal{R}_{n1}(T_{\rm CP},L)/r_{n1}$ and $\mathcal{B}_4(T_{\rm CP},L)/b_4$ as functions of $1/L$ for $L = 8$--$24~\mathrm{fm}$ at $T_{\rm CP} = 121.5348~\mathrm{MeV}$. The diamond and circle markers denote $\mathcal{R}_{n1}(T_{\rm CP},L)/r_{n1}$ and $\mathcal{B}_4(T_{\rm CP},L)/b_4$, respectively, while the dashed and dotted lines show the fit results to these values.}
    \label{fig:LYZRCP}
\end{figure}

Taking a closer look at Fig.~\ref{fig:LYZR}, however, one finds that the intersection points do not approach the limiting value (diamond marker) with increasing $L$, but their orbit deviates slightly from it. A similar behavior is also observed in ${\cal R}_{n1}(T_{\rm CP},L)$, i.e., the values at $T=T_{\rm CP}$ in all panels.

In order to understand this behavior, in Fig.~\ref{fig:LYZRCP} we plot ${\cal R}_{n1}(T_{\rm CP},L)$ and ${\cal B}_4(T_{\rm CP},L)$ normalized by their limiting values, $r_{n1}$ and $b_4$, respectively, as functions of $1/L$ down to $1/L=1/24~{\rm fm}^{-1}$. Since an accurate determination of $T_{\rm CP}$ is crucial for this analysis, we use $T_{\rm CP}=121.5348$, whose value is evaluated carefully. These ratios should approach unity for $L\to\infty$. The figure shows that ${\cal R}_{n1}(T_{\rm CP},L)/r_{n1}$ decreases and pass through ${\cal R}_{n1}(T_{\rm CP},L)/r_{n1}=1$ at $L\simeq12$~fm, but then turn to approach to unity for $L\gtrsim20$~fm.

At present, we consider that this behavior at large $L$ comes from the modification of the FSS relations due to irrelevant operators. As discussed in App.~\ref{sec:irrelevant}, the effects of the irrelevant operators alter Eqs.~\eqref{eq:Zscaling} and~\eqref{eq:fscaling}, and it gives rise to additional $L$-dependent terms in Eqs.~\eqref{eq:LYZRmix}, \eqref{eq:LYZSmix}, and~\eqref{eq:B4mix} with negative scaling exponents~\cite{Pelissetto:2000ek,Binder:2001ha,Ferrenberg:2018zst}. In our model, the term with the slowest convergence for $L\to\infty$ among them comes from $\delta\phi^5$ term in Eq.~\eqref{eq:U-exp2}, which scales as $L^{-3/4}$ at $T=T_{\rm CP}$, as discussed in Appendix~\ref{sec:irrelevant}. Since this $L$ dependence is slower than that of ${\cal R}_{nm}(T_{\rm CP},L)$ in Eq.~\eqref{eq:LYZRmix} that scales as $L^{-3/2}$, the former becomes dominant at large $L$. As discussed in Appendix~\ref{sec:irrelevant}, the contributions from higher-order terms scale as $L^{-3n/4}$ with positive integer $n$. To confirm the existence of these contributions, we have performed the three-parameter fits to ${\cal R}_{nm}(T_{\rm CP},L)$ with the fitting function
\begin{align}
    \frac{\mathcal R_{nm}(T_{\rm CP},L)}{r_{nm}} = 1 + w_1 L^{-3/4} + w_2 L^{-3/2} + w_3 L^{-9/4} \,,
    \label{eq:LYZRfit}
\end{align}
with $w_{1,2,3}$ being the fit parameters. The results of the fits for the data at $L\ge12$ are shown in Fig.~\ref{fig:LYZRCP} by the dashed lines. The figure shows that the fit reasonably describes the model results. The fit is also adopted for ${\cal B}_4(T_{\rm CP},L)/b_4$ with the same fitting function as Eq.~\eqref{eq:LYZRfit}. The result, shown by the dotted line in the figure, well describes the model result again. These results support the validity of the argument in App.~\ref{sec:irrelevant} and the importance of the irrelevant operators in the accurate determination of the location of the CP.

\section{Conclusion} \label{sec:Conclusion}

We investigated the analytic structure of the partition function in an effective model with integration over the mean-field proposed in Ref.~\cite{Kovacs:2025gct}. The integration over the spatially uniform mode provides a minimal framework for studying both the Lee–Yang zeros (LYZs) at finite volume and the Lee–Yang edge singularity (LYES) in the thermodynamic limit in a simultaneous framework. To analyze the scaling behavior near the critical point (CP), the effective potential is mapped onto the mean-field Landau potential.

We found the $T$ dependence of the LYES $\mu_{\rm ES}(T)$ in the complex $\mu_\mathrm{B}$ plane in our model is compatible with previous effective model results~\cite{Mukherjee:2021tyg, Wan:2025wdg, Zhang:2025jyv}. The LYES remains close to the real axis for $T\lesssim150$ MeV, beyond which $\Im \mu_\mathrm{ES}(T)$ rapidly increases. While $\Re \mu_\mathrm{ES}(T)$ behaves consistently with the pseudo-critical line $\mu_{\chi_2}^{\rm pc}(T)$ defined via the peak of $\chi_2$ over a wide parameter range, a large deviation is observed for $\mu_\mathrm{B}\simeq 0$.
The disagreement is modified by employing the pseudo-critical line $\mu_{\rm ES}^{\rm pc}(T)$ defined through $\Re \mu_\mathrm{ES}(T)^2$.

For fixed finite sizes, the location of the first LYZ $\mu_{\rm LY}^{(1)}(T,L)$ is significantly deviated from the LYES for $L\lesssim12$~fm. In the thermodynamic limit, they approach the LYES in the crossover and the real axis in the first-order regions, respectively. Nevertheless, for sufficiently large sizes, $L\gtrsim 8$ fm, $\Re (\mu_\mathrm{LY}^{(1)}(T,L))^2$ is insensitive to $T$. As a result, $\mu_{\rm LY}^{\rm pc}(T,L)$ defined in Eq.~\eqref{eq:mu_pcLY} has a good agreement with $\mu_{\chi_2}^{\rm pc}(T)$ in this $L$ range.  
We also studied the behavior of the LYZs with fixed $LT$ for a comparison with lattice QCD results~\cite{Basar:2023nkp, Clarke:2024ugt, Adam:2025phc}. 
In this case, the deviation of the LYZs from LYES increases at high temperatures due to the decreasing linear size. For small sizes $LT=2, 4$, even the first LYZ remains far from the LYES. Its trajectory exhibits monotonically decreasing $\Re \mu_{\rm LY}^{(1)}(T,L)$ with increasing $T$, contrary to lattice results in Ref.~\cite{Clarke:2024ugt}. 

We also investigated several methods for locating the $L\to\infty$ CP at finite sizes based on LYZs and susceptibilities~\cite{Binder:1981sa,Wada:2024qsk, Wada:2025ycz}. We tested the intersection analyses based on the Lee-Yang zero ratio (LYZR), scaled single LYZ, and Binder cumulant methods. Each quantity gives a good estimate of the CP already at moderate $L$. A detailed analysis of their size dependence leads to two main conclusions. First, scaling corrections from irrelevant operators become significant for large $L$ due to the presence of subleading odd-order contributions. This leads to a slower convergence of the intersection analysis for large sizes. Second, at very small sizes, the temperatures of the intersection points exhibit non-monotonic $L$ dependence that is not described by finite-size scaling, similarly to those observed in generalized susceptibilities at real parameters \cite{Kovacs:2025gct}. Quantities based on higher LYZs are more sensitive to these effects.

Our approach with the integration over the spatially-uniform mode can be applied to other effective models with a mean-field approach.
Its simplicity also allows for the study of additional aspects of critical behavior. For instance, it can be used for investigating the analytic structure of the partition function in the vicinity of a tricritical point~\cite{Moueddene:2024lwi}, both at finite and infinite sizes. Such an analysis would be relevant for two opposite limits of QCD: the chiral limit, in connection with the chiral tricritical point, and the heavy-quark limit~\cite{Tohme:2025nzw,Szymanski:2026dbs}, in connection with the tricritical structure of the Roberge--Weiss transition. 

Finally, the present model framework lacks several aspects, including the $L$ dependence of the potential due to the momentum discretization, deconfinement and the $\mathbb{Z}_3$ center symmetry, and hence the Roberge--Weiss transition, of QCD, which motivates a systematic improvement of our approach.
In particular, our approach can be implemented with an extended effective action, as also suggested in Ref.~\cite{Mukherjee:2021tyg}, allowing for a more comprehensive discussion of LYZs and the LYES. Restricting such an approach to a constant homogeneous background yields a constraint effective potential~\cite{Fukuda:1974ey, Fukuda:1975ib, ORaifeartaigh:1986axd}, providing a numerically tractable theory. The inclusion of nonhomogeneous classical configurations may also be important for a complete qualitative and quantitative description~\cite{Lo:2026xuc}.

\acknowledgements

We thank Chihiro Sasaki and Krzysztof Redlich for the insightful discussion.
This work was supported in part by JSPS KAKENHI Grant Numbers JP22K03619, JP24K07049, JP26KJ1395, and the Polish National Science Centre, Poland (NCN), MINIATURA 9, grant No. 2025/09/X/ST2/00748, ISHIZUE 2025 of Kyoto University, and the Center for Gravitational Physics and Quantum Information (CGPQI) at Yukawa Institute for Theoretical Physics. T.~W. is supported by Grant-in-Aid for JSPS Fellows No. 26KJ1395. G.~K. acknowledges support from the Polish National Science Centre (NCN) under OPUS Grant No. 2022/45/B/ST2/01527. 

\appendix
\section{Corrections to finite-size scaling due to irrelevant operators}
\label{sec:irrelevant}

In this appendix, we discuss the correction to the FSS relations, Eqs.~\eqref{eq:Zscaling} and~\eqref{eq:fscaling}, arising from irrelevant operators. 

In Sec.~\ref{sec:intersection}, we derived the FSS relations based on the Landau potential~\eqref{eq:Landau} that contains terms up to the fourth order. However, the potential can generally have yet higher-order terms. In order to examine their effects, in this appendix we extend Eq.~\eqref{eq:Landau} as 
\begin{align}
    U_{\rm Landau}'(\bar\phi;\tau,\eta,\{\lambda\}) 
    &= U_{\rm Landau}'(\bar\phi;\tau,\eta,\lambda_5,\lambda_6,\ldots) 
    \notag \\
    &\equiv U_{\rm Landau}(\bar\phi;\tau,\eta) + \sum_{i=5}^{N}\lambda_i\bar\phi^i
 \,,
    \label{eq:Landau'}    
\end{align}
where we assume that $N>5$ is even and $\lambda_N>0$ so that $U_{\rm Landau}'(\bar\phi;\tau,\eta,\{\lambda\})$ is bounded from below. We notice that the existence of odd-order terms are not forbidden in general systems as discussed in Sec.~\ref{sec:model}; see, Eq.~\eqref{eq:U-exp2}. From Eq.~\eqref{eq:Landau'}, the partition function 
\begin{align}
    \mathcal{Z}_{\rm Landau}'(\tau,\eta,\{\lambda\},L) 
    = \int d\bar\phi\, e^{-L^d U_{\rm Landau}'(\bar\phi;\tau,\eta,\{\lambda\})/T} \,,
    \label{eq:Z_Landau'}
\end{align}
satisfies a scaling relation 
\begin{align}
    &\mathcal{Z}_{\rm Landau}'(\tau,\eta,\{\lambda\},L) \sim \tilde{\mathcal{Z}}' (L^{y_\tau}\tau,L^{y_\eta}\eta,L^{y_5}\lambda_5,L^{y_6}\lambda_6,\ldots) \,,
    \label{eq:Zscaling'}
\end{align}
which is obtained with the scaling transformations $\lambda_i \to b^{y_i} \lambda_i$ in addition to Eq.~\eqref{eq:scale_tr} with the scaling exponents $y_i = (4-i)d/4$ and
\begin{align}
    \tilde{\mathcal{Z}}' (\tilde\tau,\tilde\eta,\{\tilde\lambda\})
    =\int d\bar\phi e^{-\tilde\tau\bar\phi^2/2-b\bar\phi^4/4!-\sum_{i=5}^N \tilde\lambda_i \bar\phi^i-\tilde\eta\bar\phi} .
\end{align}
Equation~\eqref{eq:Zscaling'} also leads to FSS of the singular part of the free energy
\begin{align}
    f_{\rm s}'(\tau,\eta,\lambda_5,\lambda_6,\cdots,L) = \tilde{f}_{\rm s}'(L^{y_\tau}\tau, L^{y_\eta}\eta, L^{y_5}\lambda_5,L^{y_6}\lambda_6,\ldots ).
    \label{eq:fscaling'}
\end{align}

Because $y_i$ for $i\ge5$ are negative, $L^{y_i}\lambda_i$ approaches zero for $L\to\infty$ in Eq.~\eqref{eq:Zscaling'}. This means that they are irrelevant operators, whose contributions are suppressed in this limit. 

Although the contribution of the irrelevant operators is suppressed for $L\to\infty$, they lead to the violation of Eqs.~\eqref{eq:Zscaling} and~\eqref{eq:fscaling} at finite $L$.
To incorporate their effects into the FSS relation, we Taylor expand Eq.~\eqref{eq:Zscaling'} and take the first order terms of $\tilde\lambda_i$ as
\begin{align}
    &\tilde{\mathcal{Z}}'(\tilde\tau,\tilde\eta,\tilde\lambda_5,\tilde\lambda_6,\ldots) 
    \notag \\
    &= \tilde{\mathcal{Z}}'(\tilde\tau,\tilde\eta,\{0\}) + \sum_{i=5}^N \tilde\lambda_i \partial_{\lambda_i} \tilde{\mathcal{Z}}' \,.
    \label{eq:Z_lambda}
\end{align}
Treating $\lambda_i$ perturbatively, one finds that Eqs.~\eqref{eq:LYZS} and~\eqref{eq:LYZR} are modified due to the second term of Eq.~\eqref{eq:Z_lambda} as~\cite{Wada:2025ycz}
\begin{align}
    &L^{y_\eta} {\rm Im} \eta_{\rm LY}^{(n)}(\tau,L) 
    \notag \\
    &= X_n + Y_n L^{y_\tau} \tau + \Lambda_{n,5} L^{y_5} + \Lambda_{n,6} L^{y_6} + \cdots \,,
    \label{eq:LYZS'} 
    \\    
    &R_{nm}(\tau,L) 
    \notag \\
    &= r_{nm} + C_{nm} L^{y_\tau} \tau + E_{nm,5} L^{y_5} + E_{nm,6} L^{y_6} + \cdots \,,
    \label{eq:LYZR'}
\end{align}
where $\Lambda_{n,i}$ and $E_{nm,i}=r_{nm} \big( \Lambda_{n,i}/X_{n} - \Lambda_{m,i}/X_{m} \big)$ are constants. 
Similarly, Eq.~\eqref{eq:B4} is modified as
\begin{align}
    & B_4(\tau,L) 
    \notag \\
    &= b_4 + C_4 L^{y_\tau} \tau + E_{4,5} L^{y_5} + E_{4,6} L^{y_6} + \cdots \,.
    \label{eq:B4'}    
\end{align}
Because $|y_5|$ is the smallest among $|y_i|$, the term containing $L^{y_5}$ has the dominant contribution in Eqs.~\eqref{eq:LYZS'}--\eqref{eq:B4'} at large $L$.

For the CP on the $T$--$\mu_\mathrm{B}$ plane, the $L$-dependence of Eqs.~\eqref{eq:LYZRmix}, \eqref{eq:LYZSmix}, and~\eqref{eq:B4mix} at $\delta T=0$ are modified by the irrelevant operators as
\begin{align}
    &R_{nm}(T_{\rm CP},L) 
    \notag \\
    &= r_{nm} \big( 1 + D_{nm} L^{2\bar y} \big) \big( 1 + E_{nm,5} L^{y_5} + E_{nm,6} L^{y_6} + \cdots \big)
    \,,
    \label{eq:LYZRmix'} \\
    &L^{y_\eta} {\rm Im} \eta_{\rm LY}^{(n)}(T_{\rm CP},L) \notag \\
    &= A_n \big( 1 + C_n L^{\bar y} \big) 
    \big( 1 + \Lambda_{n,5} L^{y_5} + \Lambda_{n,6} L^{y_6} + \cdots \big)
    \,,
    \label{eq:LYZSmix'} \\
    &B_4(T_{\rm CP},L) 
    \notag \\
    &= b_4 \big( 1 + D_4 L^{\bar y} \big) 
    \big( 1 + E_{4,5} L^{y_5} + E_{4,6} L^{y_6} + \cdots \big) \,,
    \label{eq:B4mix'}
\end{align}
with $\bar y=y_5=-3/4$ and $y_6=-3/2$ in the mean-field theories at $d=3$.

In Eq.~\eqref{eq:LYZRmix}, the leading $L$ dependence at $T=T_{\rm CP}$ is proportional to $L^{2\bar y}=L^{-3/2}$ that comes from the linear mapping~\eqref{eq:map}. On the other hand, Eq.~\eqref{eq:LYZRmix'} contains the term proportional to $L^{y_5}=L^{-3/4}$, whose convergence for $L\to\infty$ is slower than that in Eq.~\eqref{eq:LYZRmix}. Therefore, this term dominates the behavior of Eq.~\eqref{eq:LYZRmix'} at large $L$. In contrast, the leading exponent in Eqs.~\eqref{eq:LYZSmix} and~\eqref{eq:B4mix} at $T=T_{\rm CP}$ is $\bar y=-3/4$, which is the same as $y_5$. 

We emphasize that $\bar\phi^5$ term naturally emerges in the linear mapping of the CP to Landau potential, while it does not exist in the Landau potential satisfying the $Z_2$ symmetry ${\cal U}(\bar\phi)={\cal U}(-\bar\phi)$. This indicates that the leading exponent of the irrelevant operators is not universal, but differs between the systems with and without the $Z_2$ symmetry even in the same universality class. For instance, although the CP in QCD belongs to the same universality class as that of the three-dimensional Ising model, its exponent of the irrelevant contribution can be different from, and larger than, the latter's. This observation may explain the slow convergence of the intersection analyses found in Refs.~\cite{Kiyohara:2021smr,Ashikawa:2024njc,Philipsen:2021qji}.

Whereas $\lambda_5$ in Eq.~\eqref{eq:Landau'} can be nonzero, one can argue that its absolute value is small. Requiring that $U_{\rm Landau}'$ does not have two local minima at the CP, this parameter is constrained as $|\lambda_5|<4b\lambda_6$. This may explain the reason why the contribution of $L^{-3/4}$ term becomes visible only for large $L$ in Fig.~\ref{fig:LYZRCP}.

\bibliographystyle{apsrev4-1}
\bibliography{biblio}

\end{document}